\documentclass[showpacs,preprintnumbers,amsmath,amssymb]{revtex4}

\usepackage{graphicx}
\usepackage{dcolumn}
\usepackage{bm}
\usepackage{citesort}
\date{\today}
\newcommand{\bmat}{\left(\begin{array}}
\newcommand{\emat}{\end{array}\right)}
\newcommand{\be}{\begin{equation}}
\newcommand{\ee}{\end{equation}}
\newcommand{\bea}{\begin{eqnarray}}
\newcommand{\eea}{\end{eqnarray}}

\def\lsim{\raise0.3ex\hbox{$\;<$\kern-0.75em\raise-1.1ex\hbox{$\sim\;$}}}
\def\gsim{\raise0.3ex\hbox{$\;>$\kern-0.75em\raise-1.1ex\hbox{$\sim\;$}}}

\newcommand{\als}{\alpha_s}

\begin{document}
\renewcommand{\thefootnote}{\fnsymbol{footnote}}

\baselineskip 16pt
\title{Supersymmetric contributions to $ \bar{B}_s\to \phi \pi^0$ and  $
\bar{B}_s\to \phi \rho^0 $ decays in $SCET$}
\author{ Gaber Faisel}
\email{gfaisel@ncu.edu.tw}
\affiliation{ Department of Physics and Center for Mathematics and Theoretical Physics,
National Central University, Chung-li, TAIWAN 32054.}

\affiliation{Egyptian Center for Theoretical Physics, Modern
University for Information and Technology, Cairo, Egypt}

\begin{abstract}
 We study the decay modes $\bar{B}_s\to \phi
\pi^0$ and $\bar{B}_s\to \phi \rho^0$ using Soft Collinear
Effective Theory. Within Standard Model and including the error
due to the SU(3) breaking effect in the SCET parameters we find
that BR $\bar{B}_s\to \phi \pi^0 =7_{-1-2}^{+1+2}\times 10^{-8} $
and BR $\bar{B}_s\to \phi \pi^0=9 _{-1-4}^{+1+3}\times 10^{-8}$
corresponding to solution $1$ and solution $2$ of the SCET
parameters respectively.
 For the decay mode $\bar{B}_s\to \phi \rho^0$, we find that
 BR  $\bar{B}_s\to \phi \rho^0 =
20.2^{+1+9}_{-1-12}\times 10^{-8} $  and BR $ \bar{B}_s\to \phi
\rho^0 = 34.0^{+1.5 + 15}_{-1.5-22}\times 10^{-8} $ corresponding
to  solution $1$ and solution $2$ of the SCET parameters
respectively. We extend our study to include supersymmetric models
with non-universal A-terms where the dominant contributions arise
from diagrams mediated by gluino and chargino exchanges. We show
that gluino contributions can not lead to an enhancement of the
branching ratios of $\bar{B}_s\to \phi \pi^0$ and $\bar{B}_s\to
\phi \rho^0$. In addition,  we show that SUSY contributions
mediated by chargino exchange can enhance the branching ratio of
$\bar{B}_s\to \phi \pi^0$ by about 14\% with respect to the SM
prediction. For the branching ratio of $\bar{B}_s\to \phi \rho^0$,
we find that SUSY contributions can enhance its value by about 1\%
with respect to the SM prediction.

\end{abstract}
\pacs{13.25.Hw,12.60.Jv,11.30.Hv}
\maketitle
\section{ Introduction}
The decay modes $B \to K \pi  $, $ \bar{B}_s\to \phi \pi^0$ and $
\bar{B}_s\to \phi \rho^0 $ are generated at the quark level  via
$b\rightarrow s$ transition. Their amplitudes  receive
contributions from  isospin violating electroweak (EW) penguin
amplitudes. However, these contributions are expected to be small
in the case of $B \to K \pi  $ that receives large contributions
from isospin conserving QCD penguins amplitudes which are absent
in $ \bar{B}_s\to \phi \pi^0$ and $ \bar{B}_s\to \phi \rho^0 $
decays.
 Within SM, EW penguin amplitudes are small and hence the predicted
branching ratios (BR) of $ \bar{B}_s\to \phi \pi^0$ and $
\bar{B}_s\to \phi \rho^0 $ decays are so small. As a consequence,
sizeable enhancement of these branching ratios will be attributed
only  to isospin-violating new physics which can shed light on the
$K\pi$  puzzle\cite{pikpuz,Hofer:2010ee}.

Supersymmetry (SUSY) is one of the best  candidates for physics
beyond SM. SUSY provides  solution to the hierarchy problem. Moreover,
SUSY provides  new  weak CP violating phases which can
account for the baryon number asymmetry and
other CP violating phenomena in  B and K meson decays
\cite{Gabbiani:1996hi,Khalil:2001wr,Khalil:2002jq,Khalil:2006zb,Datta:2009fk,Khalil:2009zf,Faisel:2011qt}.
In addition, the effect of these phases has been studied in the CP asymmetries of
 $\tau$ decays in refs.\cite{Delepine:2006fv,Delepine:2007qg,Delepine:2008zzb}.

 The decay modes $ \bar{B}_s\to \phi \pi^0$ and $  \bar{B}_s\to \phi \rho^0 $
have been  studied within  SM in the framework of QCD
factorization in refs.\cite{Beneke:2003zv,Hofer:2010ee} and in
PQCD in ref.\cite{Ali:2007ff}. Within Soft Collinear Effective
Theory
(SCET)\cite{Bauer:2000ew,Bauer:2000yr,Chay:2003zp,Chay:2003ju},
only  the decay mode $ \bar{B}_s\to \phi \pi^0$ has been studied
in ref.\cite{Wang:2008rk}.  On the other hand, the decay modes $
\bar{B}_s\to \phi \pi^0$ and $ \bar{B}_s\to \phi \rho^0 $ have
been  studied within supersymmetry and other New Physics (NP)
beyond SM  using  QCDF in refs.~\cite{Hofer:2010ee,Hofer:2011yg}.
 The study was based on the assumption that the color-suppressed
tree-amplitude is small compared to the EW penguin amplitude and
thus any enhancement of the branching ratios of $ \bar{B}_s\to
\phi \pi^0$ and $ \bar{B}_s\to \phi \rho^0 $ can be attributed to
additional contribution to the EW penguin amplitude from NP.
However, this assumption has to be tested using a different
framework  since the color-suppressed tree-amplitude may receive
large contribution from the subleading hard spectator interaction.
Using QCDF allows us to estimate the subleading contributions from
hard spectator interaction but this estimation suffers  from large
uncertainties~\cite{Buchalla:2004tw,Hofer:2010ee}. Thus, it may be
important to analyze the relative size between the
color-suppressed tree-amplitude and the EW penguin amplitude using
a different framework such as SCET.

In this paper, we restudy the decay mode $ \bar{B}_s\to \phi
\pi^0$ using SCET to give an estimation of the error in the
predicted  branching ratio due to the SU(3) breaking effects in
the SCET parameters which is missed in the previous study. In
addition, we study the decay mode $ \bar{B}_s\to \phi \rho^0 $
using SCET and present a prediction of its  branching ratio. We
extend the study to include SUSY contributions to the branching
ratios of the $ \bar{B}_s\to \phi \pi^0$ and  $\bar{B}_s\to \phi
\rho^0 $  decays.

 SCET  is an effective field theory describing the dynamics of highly energetic particles moving close to the
light-cone interacting with a background field of soft quanta\cite{Fleming:2009fe}.
It provides a systematic and  rigorous way to deal with the
decays of the heavy hadrons that involve different energy scales. Moreover, the power counting
in SCET helps to reduce the complexity of the calculations and
 the factorization formula provided by SCET is
perturbative to all powers in $\alpha_s$ expansion.

 In SCET, we start by defining a small parameter $\lambda $
 as the ratio of the smallest to the largest energy scales in
the given process. Accordingly,  we scale  all  fields and momenta
in terms of $\lambda $. Then, the QCD lagrangian is matched into
the corresponding SCET Lagrangian which is usually  written as a
series of orders of $\lambda $. The smallness of  $\lambda $
allows us to keep terms up to order $\lambda^2 $ in the SCET
Lagrangian which in turn simplify the calculations.

We can classify two different effective theories: SCET$_{I}$ and
SCET$_{II}$ according to the momenta modes in the process under
consideration. SCET$_{I}$  is applicable in the processes in which the  momenta modes are
the  collinear and the  ultra soft  as in   the inclusive decays of a heavy meson
 such as $B\rightarrow X^*_s\gamma$ at the end point region
 and $e^-p\rightarrow e^-X$  at the threshold region in which there
 are only collinear and ultra soft  momenta modes.
  SCET$_{II}$  is applicable to the semi-inclusive or exclusive decays of a heavy
 meson such as $B\rightarrow D\pi$, $B\rightarrow K\pi$, $B\rightarrow \pi\nu e$,....etc
 in which there  are only collinear and soft momenta modes.

This paper is organized as follows. In Sec.~\ref{sec:formalism},
we briefly review the decay amplitude for $B \to M_1M_2$ within
SCET framework. Accordingly, we analyze the branching ratios of $
\bar{B}_s\to \phi \pi^0$ and  $ \bar{B}_s\to \phi \rho^0 $ decays
within SM  in section~\ref{SM}. We discuss SUSY contributions to
the branching ratios of $ \bar{B}_s\to \phi \pi^0$ and  $
\bar{B}_s\to \phi \rho^0 $ in section ~\ref{SUSY}. Finally, we
give our conclusion in Sec.~\ref{sec:conclusion}.

\section{ $ B\to {M_1M_2}$ in $ SCET $ }\label{sec:formalism}

At leading order in $\alpha_s$ expansion, the amplitude of $ B\to {M_1M_2}$ where $M_1$ and $M_2$ are light
mesons  in SCET can be written as  \bea{\cal A}_{B\to
M_1M_2}^{SCET}
 &=& {\cal A}_ {B\to M_1M_2}^{LO} +  {\cal A}_{B\to M_1M_2}^ {\chi}
 +  {\cal A}_{B\to M_1M_2}^{c.c} \eea

Here ${\cal A}_ {B\to M_1M_2}^{LO}$ denotes the leading order
amplitude in the expansion $1/m_b$, $ {\cal A}_{B\to M_1M_2}^
{\chi}$ denotes the chirally enhanced penguin amplitude generated
by corrections of order $\alpha_s(\mu_h)(\mu_M\Lambda/m_b^2)$
where $\mu_M$ is the chiral scale parameter and ${\cal A}_{B\to
M_1M_2}^{c.c}$ denotes the long distance  charm penguin
contributions. For detail discussions about the formalism of SCET
we refer to refs.\cite{Bauer:2000ew,Bauer:2000yr,Chay:2003zp,
Chay:2003ju,Jain:2007dy,Wang:2008rk}.

The  decay modes $ \bar{B}_s\to \phi \pi^0$ and $ \bar{B}_s\to \phi \rho^0 $ receive contributions
only from ${\cal A}_ {B\to M_1M_2}^{LO}$ and so we give a brief review for this amplitude
in the following.

At leading power in $(1/m_b)$ expansion, the full QCD effective
weak Hamiltonian of the $\Delta_B=1$ decays is matched into the
corresponding weak Hamiltonian in $SCET_I$ by integrating out  the
hard scale $ m_b $. Then, the $SCET_I$ weak Hamiltonian is
matched into the weak Hamiltonian $SCET_{II}$ by integrating out
the hard collinear modes with $p^2\sim \Lambda m_b$ and the
amplitude of the $\Delta_B=1$ decays  can be obtained
via~\cite{Bauer:2002aj}:

 \begin{eqnarray}
 {\cal A}^{LO}_{B\rightarrow M_1M_2}&=& - i\big\langle
M_1M_2 \big| H^{SCET_{II}}_W \big|
\bar{B}\big\rangle\nonumber\\&=&\frac{G_F
m_B^2}{\sqrt{2}}\Big(f_{M_1}\bigg[\int^1_0
 dudzT_{M_1J}(u,z)\zeta^{B M_2}_J(z)\phi_{M_1}(u)\nonumber\\
  &+& \zeta^{B M_2}\int^1_0
 du T_{M_1\zeta}(u)\phi_{M_1}(u)\bigg]+(M_1 \leftrightarrow
 M_2)\Big).\label{amp1}
 \end{eqnarray}

The hadronic parameters  $\zeta^{BM}$ and $\zeta_J^{BM}$ are
related to the form factors for heavy-to-light transitions, $B\to
M$ transitions, via the combination
$\zeta^{BM}+\zeta_J^{BM}$\cite{Bauer:2005kd}. Moreover,  Power
counting implies $\zeta^{BM}\sim \zeta^{BM}_J\sim
(\Lambda/m_b)^{3/2}$\cite{Bauer:2005kd}. Without assuming any
symmetries we would have large number of these hadronic parameters
for the 87 $B\to PP$ and $B\to VP$ decay channels. Thus it is a
common approach in SCET to use symmetries like SU(2) and SU(3)
 to reduce the number of these hadronic
parameters\cite{Bauer:2005kd,Williamson:2006hb,Wang:2008rk}. For a
model independent analysis they need to be determined from
data\cite{Jain:2007dy}. In refs.\cite{Bauer:2005kd,Jain:2007dy}
the hadronic parameters  $\zeta^{BM}$ and $\zeta_J^{BM}$ for few
decay modes of B mesons are  fitted from the experimental data. On
the other hand, in refs.\cite{Williamson:2006hb,Wang:2008rk} the
$\chi^2$ fit method using  the non leptonic decays experimental
data of  the branching fractions and CP asymmetries is used for
the determination of $\zeta^{BM}$ and  $\zeta_J^{BM}$ for large
number of $B$ and $B_s$ decays to two light mesons final states.
For details about the fit method to determine  $\zeta^{BM}$ and
$\zeta_J^{BM}$ we refer to refs.\cite{Bauer:2005kd,Jain:2007dy}.

 The hard kernels $T_{(M_1,M_2)\zeta}$ and $T_{(M_1,M_2) J}$ can be expressed in terms of
 $c_i^{(f)}$ and $b_i^{(f)}$ which are functions of the
 Wilson coefficients as \cite{Jain:2007dy}
 \begin{eqnarray}
 T_{1\zeta}(u) &=
  {\cal C}_{u_L}^{BM_2} \, {\cal C}_{f_Lu}^{M_1} \, c_1^{(f)}(u)
  +   {\cal C}_{f_L}^{BM_2} \, {\cal C}_{u_Lu}^{M_1} \, c_2^{(f)}(u)
  \nonumber \\
 & +   {\cal C}_{f_L}^{BM_2} \, {\cal C}_{u_Ru}^{M_1} \, c_3^{(f)}(u)
  +  {\cal C}_{q_L}^{BM_2} \, {\cal C}_{f_Lq}^{M_1} \, c_4^{(f)}(u)
  ,\nonumber\\
 T_{1 J}(u,z) &=
   {\cal C}_{u_L}^{BM_2} \, {\cal C}_{f_Lu}^{M_1} \, b_1^{(f)}(u,z)
  +   {\cal C}_{f_L}^{BM_2} \, {\cal C}_{u_Lu}^{M_1} \, b_2^{(f)}(u,z)
   \nonumber\\
 &+   {\cal C}_{f_L}^{BM_2} \, {\cal C}_{u_Ru}^{M_1} \, b_3^{(f)}(u,z)
  +  {\cal C}_{q_L}^{BM_2} \, {\cal C}_{f_Lq}^{M_1} \,
  b_4^{(f)}(u,z)\label{hark}
\end {eqnarray}
here $f$ stands for $d$ or $s$,  ${\cal C}_i^{BM}$ and ${\cal
C}_{i}^M$ are Clebsch-Gordan coefficients that depend on the
flavor content of the final state mesons and $c_i^{(f)}$ and
$b_i^{(f)}$ are given by \cite{Bauer:2004tj}

\begin{eqnarray}\label{cmatch}
c_{1,2}^{(f)}&=&\lambda_u^{(f)}\Big[C_{1,2}+\frac{1}{N}C_{2,1}\Big]
-\lambda_t^{(f)}\frac{3}{2}\Big[\frac{1}{N}C_{9,10}
+C_{10,9}\Big]+ \Delta c_{1,2}^{(f)},\nonumber\\
c_3^{(f)}&=&-\frac{3}{2}\lambda_t^{(f)}\Big[C_7+\frac{1}{N}C_8\Big]
+ \Delta c_3^{(f)},\nonumber\\
c_4{(f)}&=&-\lambda_t^{(f)}\Big[\frac{1}{N}C_3+C_4
-\frac{1}{2N}C_9-\frac{1}{2}C_10\Big]+ \Delta c_4^{(f)},
\end{eqnarray}
and
\begin{eqnarray}
b_{1,2}^{(f)}&=&\lambda_u^{(f)}\Big[C_{1,2}+\frac{1}{N}
\Big(1-\frac{m_b}{\omega_3}\Big)C_{2,1}\Big]-
\lambda_t^{(f)}\frac{3}{2}\Big[C_{10,9}+\frac{1}{N}
\Big(1-\frac{m_b}{\omega_3}\Big)C_{9,10}\Big]+ \Delta b_{1,2}^{(f)},\nonumber\\
b_3^{(f)}&=&-\lambda_t^{(f)}\frac{3}{2}\Big[C_7
+\Big(1-\frac{m_b}{\omega_2}\Big)\frac{1}{N}C_8\Big]+ \Delta b_3^{(f)},\nonumber\\
b_4^{(f)}&=&-\lambda_t^{(f)}\Big[C_4+\frac{1}{N}
\Big(1-\frac{m_b}{\omega_3}\Big)C_3\Big]+\lambda_t^{(f)}\frac{1}{2}
\Big[C_{10}+\frac{1}{N}\Big(1-\frac{m_b}{\omega_3}\Big)C_9\Big]+
\Delta b_4^{(f)},\label{bmatch}
\end{eqnarray}
where $\omega_2=m_bu$ and $\omega_3=-m_b\bar{u}$. $u$ and
$\bar{u}=1-u $ are the momentum fractions of the quark and
antiquark $\bar{n}$ collinear fields.  $\Delta c_i^{(f)}$ and
$\Delta b_i^{(f)}$ denote terms depending on $\alpha_s$ generated
by matching from $H_W$. The ${\cal O}(\alpha_s)$ contribution to
$\Delta c_i^{(f)}$ has been calculated in
refs.\cite{Beneke:1999br,Beneke:2000ry,Chay:2003ju} and later in
ref.\cite{Jain:2007dy} while the ${\cal O}(\alpha_s)$contribution
to $\Delta b_i^{(f)}$ has been calculated in
refs.\cite{Beneke:2005vv,Beneke:2006mk,Jain:2007dy}.  The hard
kernels  $ T_{1\zeta}(u)$ and $ T_{1 J}(u,z) $  for the decay
channel $ \bar{B}_s\to \phi \pi^0$ are given by
\cite{Williamson:2006hb}

\begin{eqnarray} \label{Tz}
  T_{1\zeta}(u) &=& 0
  \nonumber \\
 T_{2\zeta}(u) &=&
  \frac{1}{\sqrt{2}} (c_2^{(s)}(u)- c_3^{(s)}(u))
  \nonumber \\
   T_{1 J}(u,z) &=&  0  \nonumber \\
  T_{2 J}(u,z) &=&
   \frac{1}{\sqrt{2}} (b_2^{(s)}- b_3^{(s)})
\end {eqnarray}

while for $\bar{B}_s\to \phi \rho^0$,  they are given
by\cite{Williamson:2006hb}
\begin{eqnarray} \label{Tzz}
  T_{1\zeta}(u) &=& 0
  \nonumber \\
 T_{2\zeta}(u) &=&
  \frac{1}{\sqrt{2}} (c_2^{(s)}(u)+ c_3^{(s)}(u))
  \nonumber \\
   T_{1 J}(u,z) &=&  0  \nonumber \\
  T_{2 J}(u,z) &=&
   \frac{1}{\sqrt{2}} (b_2^{(s)}+ b_3^{(s)})
\end {eqnarray}

The coefficients  $ c_2^{(s)},c_3^{(s)}$ and $b_2^{(s)},b_3^{(s)}$
can be obtained from  Eqs.(\ref{cmatch}) and (\ref{bmatch})
respectively by replacing $f$ by $s$ every where. For the
definition of $\phi_{M_1}(u)$ in the case  of $\pi$ and $\rho$
mesons we use the definitions given in ref.\cite{Jain:2007dy}.

\section{ SM contributions to $ \bar{B}_s\to \phi \pi^0$ and $ \bar{B}_s\to \phi \rho^0 $  \label{SM}}

 At quark level, the decay modes $ \bar{B}_s\to \phi
\pi^0$ and $\bar{B}_s\to \phi \rho^0 $ are generated via
$b\rightarrow s $ transition. Their amplitudes receive
contributions from tree and electroweak penguin diagrams. Hence,
we can write their amplitudes ${\cal A} $ in terms of the CKM
matrix elements  as

\bea{\cal A}(\bar{B}_s\to \phi \pi^0,\phi \rho^0) &=&
\lambda_u^s{\cal A}^{tree}_u - \lambda_t^s{\cal
A}^{EW}\label{unit} \eea

Here $\lambda_p^s=V_{pb}V^*_{ps}$ with $p=u,c$ and ${\cal
A}^{tree}_u,{\cal A}^{EW}$ refer to the color suppressed tree and
electroweak penguins amplitudes respectively. Using  the unitarity
of the CKM matrix

 \be \lambda_t^s=
-\lambda_u^s-\lambda_c^s  \ee

Eq.(\ref{unit}) can be rewritten as
 \bea{\cal
A}(\bar{B}_s\to \phi \pi^0,\phi \rho^0) &=& \lambda_u^s({\cal
A}^{tree}_u+{\cal A}^{EW}_u) +\lambda_c^s{\cal A}^{EW}_c
\label{ampdeco} \eea

 where ${\cal A}^{EW}_u, {\cal A}^{EW}_c$ refer to contributions from  electroweak
 penguins which are proportional to $\lambda_u^s$ and  $\lambda_c^s$ respectively.

 In the SM, without including QCD corrections, we find that
${\cal A}^{tree}_u\gg {\cal A}^{EW}_u,{\cal A}^{EW}_c$ due to the
hierarchy of the Wilson coefficients $C_{2}\gg C_{3-10}$. Thus, we
can write, to a good approximation, \bea{\cal A}(\bar{B}_s\to \phi
\pi^0,\phi \rho^0) &\simeq& \lambda_u^s{\cal
A}^{tree}_u+\lambda_c^s{\cal A}^{EW}_c\label{ampdecoo}\eea

Eq.(\ref{ampdecoo}) indicates that  ${\cal A}^{tree}_u $ is
suppressed by a factor $|\lambda_u^s|\sim 0.02|\lambda_c^s|$
compared to ${\cal A}^{EW}_c$.  Hence, electroweak penguin
contributions to the decays $ \bar{B}_s\to \phi \pi^0$ and
$\bar{B}_s\to \phi \rho^0 $ become important and even dominant
\cite{Fleischer:1994rs}. Thus one expects that new physics
contribution to the penguin amplitude, ${\cal A}^{EW}_c$, can
enhance significantly the decay rates and accordingly the
branching ratios of these decay modes.

 Another remark here is that, since these decay modes do not receive contributions from the
long distance charm penguin, their  expected  branching ratios
will be so small as non-perturbative charming penguin plays
crucial rule in the branching ratios in SCET.

 In our analysis we take the Wilson coefficients $C_i$ at
leading logarithm order that are given by \cite{BBL}:
\bea %
C_{1-10}(m_b) &=& \{ 1.110\,,  -0.253\,, 0.011\,, -0.026\,,
0.008\,,
-0.032 \,,  0.09 \!\times\! 10^{-3} \,, 0.24  \!\times\! 10^{-3} \,,\nonumber\\
 &&  -10.3 \!\times\! 10^{-3} \,,
  2.2 \!\times\! 10^{-3} \} \,.
\label{SMW}\eea %

For the SCET parameters $\zeta^{B(M_1,M_2)}$, $\zeta_J^{B
(M_1,M_2)}$, we use the two sets of values given in
ref.\cite{Wang:2008rk}  corresponding to the two solutions
obtained from the $\chi^2$ fit method used to determine the SCET
parameters. Predictions for $B_s^0$ decays can be made using SU(3)
symmetry\cite{Williamson:2006hb,Wang:2008rk}. In our analysis, we
follow ref.\cite{Williamson:2006hb} and assume a  $20\%~$ error in
both $\zeta^{B(M_1,M_2)}$ and $\zeta_J^{B (M_1,M_2)}$  due to  the
SU(3) symmetry breaking. For the other hadronic parameters related
to the light cone distribution amplitudes we use the same input
values given in ref.\cite{Jain:2007dy}.

 After setting the input parameters and substituting  Eq.(\ref
{Tz}) in Eq.(\ref {amp1}) we  obtain the amplitude of
$\bar{B}_s\to \phi \pi^0$ decay corresponding to solution 1 of the
SCET parameters  \bea
 {\cal A}(\bar{B}^0_s\to \phi \pi^0)\times
10^6&\simeq& (-3.6 C_{10} + 1.4 \tilde{C}_{10} + 8.3 C_7 - 8.3
\tilde{C}_{7} + 1.9 C_8 - 1.9 \tilde{C}_{8} - 8.3 C_9 + 6.6
\tilde{C}_{9})\lambda^s_{t}\nonumber\\ &+& (2.4 C_1 - 0.9
\tilde{C}_{1} + 5.6 C_2 -
4.4\tilde{C}_{2})\lambda^s_{u}\label{pi1} \eea

and for solution 2 of the SCET parameters we obtain

 \bea
 {\cal A}(\bar{B}^0_s\to \phi \pi^0)\times
10^6&\simeq& (-5.1 C_{10} - 0.3 \tilde{C}_{10} + 9.3 C_7 - 9.3
\tilde{C}_{7} + 1.1 C_8 - 1.1 \tilde{C}_{8} - 9.3 C_9 + 5.2
\tilde{C}_{9})\lambda^s_{t}\nonumber\\ &+& (3.4 C_1 + 0.2
\tilde{C}_{1} + 6.2 C_2 - 3.4\tilde{C}_{2})\lambda^s_{u}
\label{pi2}\eea

Where $C_i $ and $\tilde{C}_i $ are the Wilson coefficients
which can be  expressed as
\be C_i = C_i^{SM} + C_i^{SUSY},~~~~~~~~~~~~~~~~~~~~~~~~~~~~~\tilde{C}_i = \tilde{C}_i^{SUSY} \ee

$\tilde{C}_i $ are generated from the weak effective Hamiltonian
by flipping the chirality left to right and so in the SM we have
$\tilde{C}_i^{SM}=0$. Setting $ C_i^{SUSY}=\tilde{C}_i^{SUSY}=0$
in eqs.(\ref{pi1},\ref{pi2}) and substituting the values of $
C_i^{SM}$  allow us to give our predictions for the branching
ratio of $\bar{B}^0_s\to \phi \pi^0$ which are presented in Table
\ref{branch}. As can be seen from Table \ref{branch}, our
predictions are consistent with the  SCET predictions presented in
ref.\cite{Wang:2008rk}.  Moreover, we present an estimation of the
errors due to the SU(3) symmetry breaking missed in
ref.\cite{Wang:2008rk}. Clearly from Table \ref{branch}, SCET
predictions for the branching ratios are smaller than PQCD  and
QCDF predictions. This can be explained as the predicted form
factors in SCET are  smaller than  those used in PQCD and
QCDF\cite{Wang:2008rk}.

 The uncertainties in our predictions in Table \ref{branch}
are  due to the errors in the SCET parameters $\zeta^{B(M_1,M_2)}$
and $\zeta_J^{B (M_1,M_2)}$ and the uncertainties due to the CKM
matrix elements. In other decay channels where charm penguin
contributes to their amplitudes one should add to the predictions
the errors stem from the modulus and the phase of the charm
penguin  as, in SCET,  they are also fitted from data and thus
they are given with their associated errors. After performing the
integrations in eq.(\ref{amp1}), the amplitude will be function of
$\zeta^{B(M_1,M_2)}$, $\zeta_J^{B (M_1,M_2)}$, $\lambda_u^s$ and
$\lambda_t^s$. Since we are interested to show the source of the
dominant errors in our predictions we calculate the individual
errors coming from both $\zeta^{B(M_1,M_2)}$ and  $\zeta_J^{B
(M_1,M_2)}$  and the individual errors coming from the CKM matrix
elements. In the case of calculation of the individual errors due
to $\zeta^{B(M_1,M_2)}$ and  $\zeta_J^{B (M_1,M_2)}$ we use the
central values for the CKM matrix elements and  assume a $\pm
20\%~$ errors in the central values of both $\zeta^{B(M_1,M_2)}$
and $\zeta_J^{B (M_1,M_2)}$ due to the SU(3) symmetry breaking and
thus we can proceed to calculate the corresponding errors in the
branching ratios. A similar treatment for the calculations of the
errors corresponding to the CKM matrix elements where in this case
we use the central values for  $\zeta^{B(M_1,M_2)}$ and
$\zeta_J^{B (M_1,M_2)}$ and take into account only the errors due
to the CKM matrix elements and thus we proceed to calculate the
corresponding errors in the branching ratios.

\begin{table}
\begin{center}
\begin{tabular}{|c|c|c|c|c|c|c|}
  \hline
  Decay channel & QCD factorization & PQCD & SCET solution 1 & SCET solution 2 & Prediction $1$ & Prediction $2$\\
  \hline
  $\bar{B}_s\to \phi \pi^0$  & $16_{-3}^{+11}$ & $ 16 _{-5-2-0}^{+6+2+0}$& $ 7 _{-1}^{+1}$
  &$ 9 _{-1}^{+1} $& $ 7_{-1-2}^{+1+2} $&$ 9 _{-1-4}^{+1+3}$ \\
  $\bar{B}_s\to \phi \rho^0$  & $ 44 _{-7}^{+27}$ & $23_{-7-1-1}^{+9+3+0}$&
  &  & $ 20.2^{+1+9}_{-1-12}$ & $ 34.0^{+1.5 + 15}_{-1.5-22}$  \\
  \hline
\end{tabular}
 \end{center}
\caption{ Branching ratios in units $10^{-8}$ of $ \bar{B}_s\to
\phi \pi^0$ and $\bar{B}_s\to \phi \rho^0$  decays. The last two
columns  give our predictions corresponding to the two sets
 of SCET parameters given in
Ref.\cite{Wang:2008rk}.  On our predictions we include the errors
due to  the CKM matrix elements  and  SU(3) breaking effects
respectively. For a comparison with previous studies in the
literature, we cite the results evaluated in
QCDF\cite{Hofer:2010ee}, PQCD \cite{Ali:2007ff}
 and SCET \cite{Wang:2008rk}.}\label{branch}
\end{table}

The decay mode $\bar{B}^0_s\to \phi \rho^0$ contains two vector
mesons in the final state and thus it is characterized by three
helicity
 amplitudes $A_0$ (longitudinal) and $A_{\pm}$.  Naive factorization analysis
 leads to the hierarchy  $A_0:A_-:A_+=1:\frac{\Lambda}{m_b}:(\frac{\Lambda}{m_b})^2$
where $ m_b \approx  5GeV$ is the bottom quark mass and $\Lambda
\approx 0.5GeV$ is the strong interaction
scale\cite{Korner:1979ci}. The hierarchy shows that the dominant
contribution is mainly from  the longitudinal polarization
component which has been shown  in ref.\cite{Hofer:2010ee}. Thus
in our calculation  we consider only the  longitudinal amplitude.

 After setting the input parameters and substituting  Eq.(\ref
{Tzz}) in Eq.(\ref {amp1}) we  obtain the amplitude of
$\bar{B}_s\to \phi \rho^0$ decay corresponding to solution 1 of
the SCET parameters

 \bea
 {\cal A}(\bar{B}_s\to \phi\rho^0 )\times
10^6&\simeq& (-8.3 C_{10} -4.3  \tilde{C}_{10} -11.9 C_7 +11.9
\tilde{C}_{7} + 0.4  C_8 - 0.4\tilde{C}_{8} -11.9 C_9 + 0.05
\tilde{C}_{9})\lambda^s_{t}\nonumber\\ &+& (5.5 C_1+ 2.9
 \tilde{C}_{1} + 7.9 C_2 - 0.03 \tilde{C}_{2})\lambda^s_{u}
\label{rho1}\eea

and for solution $2$ of the SCET parameters  we obtain

 \bea
 {\cal A}(\bar{B}_s\to \phi\rho^0 )\times
10^6&\simeq& (-7.4 C_{10} +0.33  \tilde{C}_{10} -14.9 C_7 +14.9
\tilde{C}_{7} - 2.5  C_8 + 2.5 \tilde{C}_{8} - 14.9  C_9 + 8.3
\tilde{C}_{9})\lambda^s_{t}\nonumber\\ &+& (4.9 C_1-
 0.22  \tilde{C}_{1} + 9.9 C_2 -5.5 \tilde{C}_{2})\lambda^s_{u}
\label{rho2}\eea

As before, we set $ C_i^{SUSY}=\tilde{C}_i^{SUSY}=0$ in
Eqs.(\ref{rho1},\ref{rho2}) then we substitute the values of $
C_i^{SM}$ and we proceed to obtain the predictions for the
branching ratios of $\bar{B}^0_s\to \phi \rho^0$ within SCET which
are presented in Table \ref{branch}.  These results account for
the SCET prediction of the branching ratio of $\bar{B}^0_s\to \phi
\rho^0$ for the first time. Clearly from Table \ref{branch}, SCET
predictions for the branching ratio of $\bar{B}^0_s\to \phi
\rho^0$ are smaller than  QCDF predictions for the same reason
mentioned above in the case of $\bar{B}^0_s\to \phi \pi^0$.

 As can be seen from Table \ref{branch}, the
branching ratios of $\bar{B}^0_s\to \phi \rho^0$ are larger  than
the branching ratios of $\bar{B}^0_s\to \phi \pi^0$ in agreement
with the QCDF prediction in ref.\cite{Hofer:2010ee}. The two
decays $\bar{B}^0_s\to \phi \rho^0$ and $\bar{B}^0_s\to \phi
\pi^0$ are generated via the $\bar{B}_s \to \phi$ transition and
thus they have the same non perturbative form factors
$\zeta^{B\phi}$ and $\zeta_J^{B\phi}$. However, using a
non-polynomial model for the light cone distribution amplitude
$\phi_{\rho}(u)$ can lead to a slightly different result from
using the polynomial model for the light cone distribution
amplitude in the case of $\phi_{\pi}(u)$ as pointed out in
ref.\cite{Jain:2007dy}. Another reason for this difference is due
to the opposite sign for the coefficients $c^s_3 (u) $ and $
b^s_3$ in the  hard kernels, $ T_{1\zeta}(u)$ and $ T_{1 J}(u,z)
$,  as can be seen from Eq.(\ref{Tz}) and Eq.(\ref{Tzz}).

 In the SCET formalism, the hard
kernels $ T_{i J}(u,z) $ where $ i=1,2 $  account for the
subleading hard spectator interaction. The non-vanishing values of
$T_{2 J}$ in Eqs.(\ref{Tz},\ref{Tzz}) show that the amplitudes of
$ \bar{B}_s\to \phi \pi^0$ and $\bar{B}_s\to \phi \rho^0 $ receive
contributions from the hard spectator interaction. Denoting the
hard spectator interaction contributions to the color-suppressed
tree-amplitude by ${\mathcal C}^H$ and using Eq.(\ref{amp1}) we
find that

 \begin{eqnarray} {\mathcal C^H} &=& \frac{G_F
m_B^2}{\sqrt{2}}\Big(f_{M_1} \int^1_0 dudzT'_{M_1J}(u,z)\zeta^{B
M_2}_J(z)\phi_{M_1}(u) +(M_1 \leftrightarrow
M_2)\Big).\label{CHan1}
 \end{eqnarray}

 where the hard kernels $ T'_{i J}(u,z) $, for $ i=1,2 $, can be obtained from
 $ T_{i J}(u,z) $ by setting  $\lambda^s_t=0$ in the coefficients
$b_{2,3}^{(s)}$. After setting the input parameters in
Eq.(\ref{CHan1}), we find that \be {\mathcal C^H}  =
(1.2^{+0.2}_{-0.3}+2.7^{+0.5}_{-0.6}\, i)\times 10^{-9}, \ee
corresponding to solution 1 of the SCET parameters for the decay
$\bar{B}_s\to \phi \rho^0 $ while for solution 2 of the SCET
parameters we find that \be {\mathcal C^H} =
(0.7^{+0.1}_{-0.2}+1.5^{+ 0.3}_{-0.3}\, i)\times 10^{-9} ,\ee

Turning now to the decay mode $ \bar{B}_s\to \phi \pi^0$ we find
that
 \be
{\mathcal C^H} =  (1.0^{+0.2}_{-0.2} +2.1^{+0.4}_{-0.4} \,
i)\times 10^{-9},\ee corresponding to solution 1 of the SCET
parameters while for solution 2 of the SCET parameters we find
that

\be {\mathcal C^H} = (0.5^{+0.1}_{-0.1}+1.2^{+0.2}_{-0.2}\,
i)\times 10^{-9}\ee

The uncertainties in the predictions for ${\mathcal C}^H$ are
mainly due to the errors in the SCET parameter $\zeta^{BM}_J$
where we assume a  $20\%~$ error due to  the SU(3) symmetry
breaking as we referred to in the beginning of this section. The
other uncertainties are due to the CKM matrix elements are much
less important and so we did not take them into account in the
predictions of ${\mathcal C}^H$.  Comparing the results of
${\mathcal C^H}$ for both solutions 1 and 2 of the SCET
parameters, for both $\bar{B}_s\to \phi \rho^0 $ and $\bar{B}_s\to
\phi \pi^0 $ decays, show that solution 1 leads to a larger
${\mathcal C}^H$ than what solution 2 can lead to. The reason is
that the non- perturbative parameter $\zeta^{BM}_J$ enters in the
calculation of the hard spectator interaction has two different
values from the fit and the ratio of $\zeta^{BM}_J$ corresponding
to solution 1 to  $\zeta^{BM}_J$ corresponding to solution 2
$\simeq 1.8$.

 Turning now to the evaluation of  the total color-suppressed
tree-amplitude (${\mathcal C}$) that can be expressed using
Eq.(\ref{amp1}) as
 \begin{eqnarray}
 {\mathcal C} &=&\frac{G_F
m_B^2}{\sqrt{2}}\Big(f_{M_1}\bigg[\int^1_0
 dudzT'_{M_1J}(u,z)\zeta^{B M_2}_J(z)\phi_{M_1}(u)\nonumber\\
  &+& \zeta^{B M_2}\int^1_0
 du T'_{M_1\zeta}(u)\phi_{M_1}(u)\bigg]+(M_1 \leftrightarrow
 M_2)\Big).\label{Ctotal}
 \end{eqnarray}
   The hard kernels $T'_{M_1\zeta}(u)$ and
 $ T'_{i J}(u,z) $, for $ i=1,2 $, can be obtained from
 $T_{M_1\zeta}(u)$ and $ T_{i J}(u,z)$ by setting
 $\lambda^s_t=0$ in the coefficients $c_{2,3}^{(s)}$
 and $b_{2,3}^{(s)}$ respectively. Upon substituting the input parameters in Eq.(\ref{Ctotal}),
  we find that in the case of $\bar{B}_s\to \phi \rho^0 $ decay  \be {\mathcal C}
= (-4.2^{+0.8}_{-0.8}-0.08^{+0.02}_{-0.01}\, i)\times 10^{-9} \ee
corresponding to solution 1 of the SCET parameters while for
solution 2 of the SCET parameters we get \be {\mathcal C} =
(-5.5^{+1.0}_{-1.0}-0.1^{+ 0.02}_{-0.02}\, i)\times 10^{-9} \ee

For the decay mode $ \bar{B}_s\to \phi \pi^0$ we find that
 \be
{\mathcal C} =  (-3.2^{+0.6}_{-0.6} -0.06^{+0.01}_{-0.01} \,
i)\times 10^{-9}\ee corresponding to solution 1 of the SCET
parameters while for solution 2 of the SCET parameters we find
that

\be {\mathcal C} = (-3.5^{+0.7}_{-0.7}-0.06^{+0.01}_{-0.01}\,
i)\times 10^{-9}\ee We see from the results for ${\mathcal C}$ in
both decays that the maximum value of the uncertainty can be about
20\%. The sources of uncertainties in ${\mathcal C}$  are due to
the non-perturbative parameters $\zeta^{B(M_1,M_2)}$ and
$\zeta_J^{B (M_1,M_2)}$ and the CKM elements. As before, the
largest uncertainties are due to the errors in the
non-perturbative parameters $\zeta^{B(M_1,M_2)}$ and $\zeta_J^{B
(M_1,M_2)}$ and thus we neglect the  uncertainties due to the CKM
elements. Another remark about the relative size of hard spectator
interaction to the total color-suppressed tree-amplitude can be
noticed by  comparing the results of ${\mathcal C^H}$ and
${\mathcal C}$. These results  show that the color-suppressed
tree-amplitude can receive large contribution from the hard
spectator interaction only for the case corresponding to solution
1 of the SCET parameters. The reason is, as explained above, due
to the value of $\zeta^{BM}_J$ corresponding to solution 1 is
$\simeq 1.8$ larger than that of  $\zeta^{BM}_J$ corresponding to
solution 2.

As we have shown above, using  SCET framework,  we can predict the
value of the total color-suppressed tree-amplitude  with
uncertainties up to 20\%. Moreover we can predict the contribution
from the hard spectator interaction to the total color-suppressed
tree-amplitude with uncertainties up to 20\% also. This is somehow
similar to  QCDF  where the color-suppressed tree-amplitude
suffers from large spectator-scattering uncertainties due to a
strong cancellation between the leading order and QCD vertex
corrections\cite{Hofer:2010ee}.

 Finally, we compare the total color-suppressed tree- amplitude with the total
electroweak penguin amplitude. For this comparison, we define the
ratio $ R = \frac {|\lambda_u^s{\cal A}^{tree}_u
|}{|\lambda_t^s{\cal A}^{EW}|}$ which gives the relative size of
the color-suppressed tree- amplitude compared to the electroweak
penguin amplitude. We find that $ R \simeq  0.32\pm 0.13 $ and $ R
\simeq 0.51 \pm 0.20 $ for the amplitudes of $ \bar{B}_s\to \phi
\pi^0$ given in Eq.(\ref{pi1}) and Eq.(\ref{pi2}) respectively and
the uncertainties in $R$ are due to the SU(3) symmetry breaking
effects as before.   For the decay $\bar{B}_s\to \phi \rho^0$ we
find that $ R \simeq 0.79 \pm 0.32 $ and $ R \simeq 0.43 \pm 0.17
$ for the amplitudes given in Eq.(\ref{rho1}) and Eq.(\ref{rho2})
respectively. The results show that the electroweak penguin
amplitude is dominant in both decays in agreement with the QCDF
results in refs.\cite{Fleischer:1994rs,Hofer:2010ee}. As a
consequence, it is suitable to look for NP in these decay modes as
additional contributions to the electroweak penguin amplitudes
from NP can enhance their  branching ratios sizably and thus
making them observable at LHCb or Super-B factory.

 In the next section, we analyze  SUSY contributions to the
branching ratios of $ \bar{B}_s\to \phi \pi^0$ and $\bar{B}_s\to
\phi \rho^0$ decays.

\section{ SUSY contributions to the
branching ratios of $ \bar{B}_s\to \phi \pi^0$ and $\bar{B}_s\to
\phi \rho^0$ decays \label{SUSY}}

New Physics  contributions  to the Wilson coefficients of
 $\bar{B}_s\to \phi \pi^0$ and $\bar{B}_s\to \phi
\rho^0$ may lead to an  enhancement of  their  branching ratios.
This possibility has been studied in ref.\cite{Hofer:2010ee}
within QCD factorization for many models beyond SM including
supersymmetry. In their study,  the authors adopted exact
diagonalization of squark mass matrices  and found that the
enhancements in the Wilson coefficients due to SUSY contributions
are not sufficient to enhance the branching ratios of $
\bar{B}_s\to \phi \pi^0$ and $ \bar{B}_s\to \phi \rho^0 $
sizeably. In this section we  check this finding within SCET and
adopt also the exact diagonalization of squark mass matrices in
our analysis.

Throughout this section, we use the MSSM convention of
ref.\cite{Cho:1996we} and diagonalize the squark mass matrices
exactly by the two $6\times 6$  matrixes $ \Gamma^{U}$ and $
\Gamma^{D}$. The $6\times 3$ block components of $\Gamma^{U}$ and
$\Gamma^{D}$ are defined via

\bea \Gamma^{U}_{6\times 6}&=&(\Gamma^{U_L}_{6\times
3}\,\,\,\,\Gamma^{U_R}_{6\times
3})\nonumber\\
\Gamma^{D}_{6\times 6}&=&(\Gamma^{D_L}_{6\times
3}\,\,\,\,\Gamma^{D_R}_{6\times 3}) \eea

The dominant SUSY contributions to the Wilson coefficients come
from diagrams with  gluino and chargino exchanges  and so we can
write

\be C_i^{SUSY} = C_i^{\tilde{g}} + C_i^{\tilde{\chi}}, \ee where
$C_i^{\tilde{g}}$ represents the gluino contribution and
$C_i^{\tilde{\chi}}$ represents the chargino contribution. The
relevant diagrams for Wilson coefficients of our processes can be
found in ref.\cite{Cho:1996we} with replacing the lepton pair by
quark pair and sneutrino by squarks. The expressions for gluino
and chargino contributions to the Wilson coefficients in terms of
these $6\times 3$  block components are listed in the Appendix.

 In Ref.\cite{Huitu:2009st}, it was pointed out that
 gluino-mediated photon penguin diagrams can lead to a significant amount of
Isospin-violation sufficient to explain the $\Delta A_{CP}(B \to
\pi K)$ data. However, this possibility is not true as  there is a
missing factor $-\alpha/6\pi $ in $C_{9g}$ used in
ref.\cite{Huitu:2009st} as pointed out in Ref.\cite{Hofer:2010ee}.
A recent analysis  of gluino-mediated photon penguin contributions
to the isospin-violation has been carried in
ref.\cite{Hofer:2010ee}. Their analysis shows that, the
contributions from gluino-mediated photon penguin are below the
3\% level to the SM coefficients of the EW penguin operators.  As
a consequence, their  conclusion is that no sizeable enhancement
of EW penguins in the MSSM with flavour-violation in the
down-sector. In our analysis we  keep all gluino contributions to
the  Wilson coefficients in order to get a clear conclusion about
their effect on the Isospin-violation.

In order to evaluate numerically SUSY contributions to the Wilson
coefficients we need to specify explicit values for the parameters
in the superpotential and the soft supersymmetric breaking
Lagrangian. In
refs.\cite{Masiero:1999ub,Abel:1996eb,Khalil:1999zn,Khalil:1999ym,Barbieri:1999ax,Babu:1999xf,Brhlik:1999hs,Bailin:2000ev,Khalil:2000ci,Kobayashi:2000br,Everett:2001yy},
it has been shown that non universality of the trilinear
interaction couplings $A^{U,D,E}$ is very relevant in the low
energy observables. Thus, in our analysis we assume  non
universality of these couplings in the quark sector only for
simplicity and write them
 as  \be A^{U,D}=
\tilde{A}^{U,D}\cdot Y^{U,D}\ee  and for the lepton sector we
assume \be A^{E}_{ij}= A_0Y^{E}_{ij}\ee where $Y^{U,D,E}$ denote
the the fermions Yukawa matrices and for reducing the number of
the free parameters we assume that $A_0$ is real. The matrices
$\tilde{A}^{U,D}$ have in total 18 complex free parameters. In
ref.\cite{Kobayashi:2000br} it has been shown that in generic
models of SUSY breaking these 18 complex free parameters can be
reduced to 9 complex parameters for $\tilde{A}^{U}$ and
$\tilde{A}^{D}$. Moreover, the magnitudes of these parameters are
order of the gaugino and soft scalar
masses\cite{Kobayashi:2000br}. In general $\tilde{A}^{U}$ and
$\tilde{A}^{D}$ can  have different structure but for simplicity
we assume that $\tilde{A}^{U} = \tilde{A}^{D} $. In addition, we
follow ref.\cite{Bailin:2000ev} and  parameterize $ \tilde{A}^{U}
$ and $ \tilde{A}^{D} $  as \be
 \tilde{A}^{U} = \tilde{A}^{D} = \left (
\begin{array}{ccc}
a & a & b\\
a & a & b \\
b & b & c
\end{array} \right).
\label{Atex} \ee

 where the entries $a,b$ and $c$ are complex and of
order the gaugino and soft scalar masses.  After rotating the
phase of the gaugino masses the $\tilde{A}$--sector will have
three phases: $\phi_a$, $\phi_b$ and $\phi_c$, which are the
relative phases between the gaugino phase and the original phases
of the $\tilde{A}_{ij}$ entries\cite{Bailin:2000ev}. In our
analysis  we  apply the constraints imposed on $A^{U,D,E}$ from
the vacuum stability argument regarding the absence in the
potential of color and charge breaking minima and of directions
unbounded from below \cite{Casas:1995pd,Casas:1996de}. Moreover,
we apply the constraints from the electric dipole moments (EDM)
\cite{Abel:1996eb}. The limits from the EDM of the electron and
the neutron constrain $\phi_a$ while the limits from the EDM of
the mercury atom constrain  $\phi_c$ and so we set $\phi_a =\phi_c
= 0$ \cite{Bailin:2000ev}. Thus, the free parameters we need are
 $\vert a \vert $, $\vert b \vert$, $\vert c \vert$, $m_0$,
$m_{1/2}$, $A_0$, $\phi_b$, $\tan\beta$ and all Standard Model
fermion and gauge boson masses and couplings. Here $m_0$ and
$m_{1/2}$ denote the common soft scalars and gaugino masses
respectively.

  The $\mu$ and $B$ parameters can be determined from
tree level relations \cite{Cho:1996we}
\bea  |\mu|^2 &=& \frac{m_2^2 \sin^2\beta - m_1^2 \cos^2\beta }{
\cos 2 \beta}
        - \frac{1}{2} m_Z^2 \nonumber\\
B \mu &=& \frac{1}{2} \sin 2 \beta(m_1^2 + m_2^2 + 2 |\mu|^2) \eea

Where $m_1$ and $m_2$ are the scalar mass-squared terms for the
higgs. The phase of the $\mu$ parameter is tightly constrained by
neutron electric dipole moment limits \cite{Buchmler} and so we
shall simply take $\mu$ to be real. In our analysis we perform a
scan over the MSSM parameter space in the following ranges

\be 2\leq\tan \beta\leq 55,\,\,\,\, 400 \leq m_{1/2} \leq
1400,\,\,\,\, 300 \leq m_0 \leq 1400,\,\,\,50 \leq m_{A} \leq
2700,\,\,\,\,-\pi \leq \phi_b \leq \pi,\,\ee

For the other parameters, we set $\vert a \vert = 3 m_0 $, $\vert
b \vert= 2 m_0 $, $\vert c \vert = m_0 $ and  $A_0 = m_0$ to
reduce the number of the free parameters in our scan. The range of
$\tan\beta $  ensures that Landau poles do not develop in the top
or bottom Yukawa couplings anywhere between the weak and GUT
scales \cite{Cho:1996we}.

After the scalar, gaugino and Yukawa terms in the soft
supersymmetry breaking sector are evaluated at $M_{GUT}\sim
10^{16} GeV$ and run down to $\mu=m_Z$  using the RGE  listed in
appendix A of ref \cite{Bertolini:1990if}, the numerical values of
the MSSM parameter space can be determined. We should take into
account all relevant constraints imposed on this parameter space.
We  reject all points in the MSSM parameter space which yield
negative values for $|\mu|^2$ or $B \mu$ as these points fail to
break the electroweak symmetry \cite{Cho:1996we}. In addition, the
following two  conditions
\bea   |B \mu|^2>  (m_1^2  &+&| \mu|^2)(m_2^2 + |\mu|^2)
\nonumber\\  (m_1^2-m_2^2) \cos 2 \beta &>& 0 \eea
must  be satisfied in order to have a stable scalar potential
minimum\cite{Cho:1996we}.

We apply also the constraints from direct search for SUSY
particles in colliders. The ATLAS and CMS collaborations search
for the superparticles at the Large Hadron Collider (LHC) provide
stringent limits on the masses of colored superparticles
\cite{Aad:2011ib,Chatrchyan:2011zy}. Barring accidental features
such as spectrum degeneracies, gluinos $\tilde{g}$ and squarks of
the first two generations  have been ruled out for masses up to
about 1 TeV \cite{Aad:2011ib,Chatrchyan:2011zy,Berger:2011af}. On
the other hand the LHC bounds on third-generation squarks are
quite weak: stops above 200-300 GeV are currently allowed.  At
present, gluinos above 600 GeV are allowed if decaying only via
the 3rd generation\cite{Berger:2011af}. For the masses of the
chargino and sparticle of the lepton sector, we take into account
the bounds from the LEP direct search\cite{pdgrev}.

  We now discuss other important constraints that should be
taken into account. We start by considering the constraints from $
B\to X_q\gamma $ decays where q refers to $d$ or $s$ quark. The
decay mode $ B\to X_s\gamma $ has been studied in the literature
in refs.
\cite{D'Ambrosio:2002ex,Bertolini:1990if,Buras:2002vd,Carena:2000uj,Bertolini:1986tg,Bobeth:1999ww,Ciuchini:1998xy,Borzumati:2003rr,Degrassi:2006eh,Degrassi:2000qf}
and in refs.
\cite{Gabbiani:1996hi,Gabbiani:1988rb,Hagelin:1992tc,Borzumati:1999qt,Besmer:2001cj,Ciuchini:2002uv,Gabrielli:2000hz,kagan,Khalil:2005qg}.
On the other hand the decay mode $ B\to X_d\gamma $ has been
studied within SM in
refs.~\cite{Ali:1992qs,Ricciardi:1995jh,Ali:1998rr} and within
supersymmetry in refs.
\cite{Akeroyd:2001cy,Hurth:2003dk,Crivellin:2011ba}. In the SM,
the general effective hamiltonian  governing  $\bar B\to X_q
\gamma$ decays  is given by \cite{Hurth:2003dk}

\begin{equation}
H_{\rm eff} (b\rightarrow q \gamma) = - {4 G_F\over \sqrt{2}}
V_{tb}^{} V_{tq}^* \left( \sum_{i=1}^8 C_i (\mu)\cdot O_i (\mu) +
\epsilon_q \, \sum_{i=1}^2 C_i (\mu)\cdot (O_i (\mu)-O_i^u (\mu)
)\right), \label{bqgEH}
\end{equation}
where $\epsilon_q = (V_{ub}^{} V_{uq}^*) / (V_{tb}^{} V_{tq}^*)$
and the four quark operators are: \\
\noindent
\begin{minipage}{0.45\linewidth}
\begin{eqnarray}
\hskip -0.5cm && O_1^u = (\bar{q}_{L} \gamma_\mu T^a u_{ L}) (\bar{u}_{ L} \gamma^\mu T^a b_{ L}), \nonumber \\
\hskip -0.5cm && O_1  = (\bar{q}_{L} \gamma_\mu T^a c_{ L}) (\bar{c}_{ L} \gamma^\mu T^a b_{ L}),\nonumber  \\
\hskip -0.5cm && O_3 = (\bar{q}_{L} \gamma_\mu b_{ L}) \textstyle \sum_{q'} ({\bar{q}'}_{ L} \gamma^\mu {q'}_{ L}),\nonumber \\
\hskip -0.5cm && O_5 = (\bar{q}_{L} \gamma_\mu \gamma_\nu
\gamma_\rho b_{ L}) \textstyle \sum_{q'} ({\bar{q}'}_{ L}
         \gamma^\mu \gamma^\nu \gamma^\rho {q'}_{ L}),\nonumber \\
\hskip -0.5cm && O_7 =\frac{e }{16 \pi^2} m_b (\mu) (\bar{q}_{L}
\sigma_{\mu \nu} b_{R}) F^{\mu \nu},\nonumber
\end{eqnarray}
\end{minipage}
\begin{minipage}{0.51\linewidth}
\begin{eqnarray}
\hskip -0.5cm && O_2^u = (\bar{q}_{L} \gamma_\mu u_{ L}) (\bar{u}_{ L} \gamma^\mu b_{ L}),\nonumber\\
\hskip -0.5cm && O_2 = (\bar{q}_{L} \gamma_\mu c_{ L}) (\bar{c}_{ L} \gamma^\mu b_{ L})\nonumber\\
\hskip -0.5cm && O_4 = (\bar{q}_{L} \gamma_\mu T^a b_{ L}) \textstyle \sum_{q'} ({\bar{q}'}_{ L} \gamma^\mu T^a {q'}_{ L}),\label{operatorbasis} \\
\hskip -0.5cm && O_6 = (\bar{q}_{L} \gamma_\mu \gamma_\nu
\gamma_\rho T^a b_{ L}) \textstyle \sum_{q'} ({\bar{q}'}_{ L}
         \gamma^\mu \gamma^\nu \gamma^\rho T^a {q'}_{ L}),\nonumber\\
\hskip -0.5cm && O_8  = \frac{g_s }{16 \pi^2} m_b (\mu)
(\bar{q}_{L} T^a \sigma_{\mu \nu} b_{R }) G^{a \mu \nu} \; .
\nonumber
\end{eqnarray}
\end{minipage}

Within supersymmetry, the dominant effects only modify the Wilson
coefficients of the dipole operators $O_7$ and $O_8$.  In
addition, SUSY has new contributions to the Wilson coefficients of
the dipole operators with opposite chirality:
\begin{equation}
O_7^R = \frac{e }{16 \pi^2} m_b (\mu) (\bar{q}_{R} \sigma_{\mu
\nu} b_{L}) F^{\mu \nu}\,,  \quad  \quad O_8^R = \frac{g_s }{16
\pi^2} m_b (\mu) (\bar{q}_{R} T^a \sigma_{\mu \nu} b_{L}) G^{a \mu
\nu} \; . \label{operatorchiral}
\end{equation}

In our analysis we consider only the  sizeable  contribution to
the branching ratio of $ B\to X_q\gamma $ and hence we use the NLO
formula of ref.\cite{Hurth:2003dk}

 \bea Br [\bar B \to X_q \gamma] & = & {{\cal N}\over 100} \,
\left| V_{tq}^* V_{tb}^{}\over V_{cb}^{}\right|^2  \, \Big[
a  + a_{77} \, (|R_7|^2 + |\widetilde R_7|^2)+ a_7^r \, {\rm Re} (R_7) + a_7^i \, {\rm Im} (R_7) \nonumber \\
& & \hskip -1.9cm + a_{88} \, (|R_8|^2+ |\widetilde R_8|^2) +
a_8^r \, {\rm Re} (R_8) + a_8^i \, {\rm Im} (R_8) +
a_{\epsilon\epsilon} \, |\epsilon_q|^2 + a_\epsilon^r \, {\rm Re}
(\epsilon_q) + a_\epsilon^i \, {\rm Im} (\epsilon_q)
\nonumber \\
 & &  \hskip -1.9cm
+ a_{87}^r \, {\rm Re} (R_8^{} R_7^* + \widetilde R_8^{}
\widetilde R_7^*) + a_{7\epsilon}^r \, {\rm Re} (R_7^{}
\epsilon_q^*) +
      a_{8\epsilon}^r\, {\rm Re} (R_8^{} \epsilon_q^*)
\nonumber \\
& &  \hskip -1.9cm + a_{87}^i \, {\rm Im} (R_8^{} R_7^*+
\widetilde R_8^{} \widetilde R_7^*) + a_{7\epsilon}^i \, {\rm Im}
(R_7^{} \epsilon_q^*) + a_{8\epsilon}^i \, {\rm Im} (R_8^{}
\epsilon_q^*) \Big] \, , \label{brnum} \eea
here ${\cal N}=2.567 \, (1 \pm 0.064 ) \times 10^{-3}$ and the
numerical values of the coefficients  introduced in
eq.(\ref{brnum}) can be found in Table 1 in
ref.\cite{Hurth:2003dk}. The CP conjugate branching ratio, $Br [B
\to X_q \gamma]$, can be obtained by Eq.~(\ref{brnum}) by
replacing ${\rm Im} (...) \rightarrow -{\rm Im} (...) $. In
Eq.~(\ref{brnum}) the ratios $R_{7,8}$ and $\widetilde R_{7,8}$
are defined as
\bea R_{7,8} = {C_{7,8}^{(0) {\rm tot}}(\mu_0) \over C_{7,8}^{(0)
{\rm SM}}(m_t)} \quad \hbox{and} \quad \widetilde R_{7,8} =
{C_{7R,8R}^{(0) {\rm SUSY}}(\mu_0) \over C_{7,8}^{(0) {\rm
SM}}(m_t)} \eea

The values of $ {C_{7,8}^{(0) {\rm SM}}(m_t)} $ can be found in
ref.\cite{Hurth:2003dk}. $C_{7,8}^{(0){\rm tot}} $ and
${C_{7R,8R}^{(0) {SUSY}}}$  receive large contributions from
diagrams mediated by gluino and down squarks exchange and diagrams
with chargino and up squarks exchange. We refer to SUSY
contributions to  $C_{7,8}^{(0){\rm tot}} $ and ${C_{7R,8R}^{(0)
{SUSY}}}$ in the following  by $C_{7\gamma,8g}$ and
$\widetilde{C}_{7\gamma,8g}$ respectively. In
ref.\cite{Crivellin:2008mq}, the dominant supersymmetric radiative
corrections to the couplings of charged Higgs bosons and charginos
to quarks and squarks are derived in the Super-CKM basis. On the
other hand, in refs.\cite{Crivellin:2008mq, Crivellin:2009ar} it
was pointed out that chirally enhanced supersymmetric QCD
corrections arising from flavor-changing self-energy diagrams can
numerically dominate over the leading-order one-loop diagrams. The
complete resummation of the leading chirally-enhanced corrections
stemming from gluino-squark, chargino-sfermion and
neutralino-sfermion loops in the MSSM with non-minimal sources of
flavor-violation can be found in ref.\cite{Crivellin:2011jt}. In
the decoupling limit $M_{susy}\gg v$, all these leading
chirally-enhanced corrections can be included into perturbative
calculations of Feynman amplitudes
\cite{Crivellin:2008mq,Crivellin:2009ar,Crivellin:2011jt}. For
large value of $|\mu| \tan \beta$, chirally enhanced
supersymmetric QCD corrections are large for heavy  squarks and
gluino\cite{Crivellin:2009ar}.  Thus taking into account these
corrections can lead to strong constraints on the SUSY parameter
space \cite{Crivellin:2009ar,Crivellin:2011ba}.  For the sake of
simplicity, we only list  the expressions for supersymmetric
contributions at leading order for all processes under
consideration. The inclusion of the chirally enhanced
supersymmetric QCD corrections into the  calculations can be
simply achieved via the procedure presented in
ref.\cite{Crivellin:2009ar}

 In the Appendix we list the leading order calculations of gluino
and chargino contributions to the Wilson coefficients
$C_{7\gamma,8g}$ and $\widetilde{C}_{7\gamma,8g}$ relevant to
$b\to s\gamma$ from which we can easily obtain the contributions
to $b\to d\gamma$. Taking  into account the NNLO correction to
${\rm BR}^{\rm{SM}}(B\to X_s\gamma)$ \cite{Misiak:2006zs} and
including the experimental errors  and the theoretical
uncertainties ~\cite{bphys,Abazov:2010fs} we  obtain the following
bound $ 2.77\times 10^{-4} \leq {\rm BR}(B\to X_s\gamma)  \leq
4.37\times  10^{-4}$. For the $b\to d\gamma$, the new NLO SM
prediction is $\langle Br \left[ B\to X_d\gamma
\right]^{SM}_{E_{\gamma}>1.6 {\rm GeV}}\rangle =
\;1.54^{+0.26}_{-0.31}\times10^{-5}$  where $\langle...\rangle$
denotes CP averaging \cite{Crivellin:2011ba}. This prediction is
well within the experimental $1\sigma$ range as  we have $\langle
{Br}\left[ B \to X_d \gamma \right]^{exp}_{E_{\gamma}>1.6 {\rm
GeV}}\rangle=(1.41\pm0.57) \times 10^{-5}$
\cite{:2010ps,Wang:2011sn}.  Thus, for  constraining our parameter
space we  require that branching ratio, including SUSY
contributions, should lie within the $2\sigma$ range of the
experimental values.

Next we consider the constraints from ${B}_q-\bar{B}_q $ mixing
where $q=d,s$. Within SM, the mass difference between the neutral
${B}_q$ states, $\Delta M^{SM}_{B_q}$, is given by \cite{BBL}
 \be
\Delta M^{SM}_{B_q}=\frac{G^{2}_{F} m^2_{W}}{12 \pi^2} \eta_{B}
m_{B_q}f^{2}_{B_q}\hat{B}_q ( V^*_{tq}V_{tb})^2 S_{0}(x_t)
 \ee
with $S_{0}(x_t)=0.784 x_t^{0.76}$, $x_{t}=(m_t/m_W)^2$,
$\eta_{B}= 0.552$ is the QCD correction to $S_0(x_t)$. The
non-perturbative hadronic parameters $\hat{B}_q $ and $ f_{B_q}$
are the bag parameters and decay constant respectively. The
supersymmetric contributions to $\Delta M_{B_q}$ in mass
eigenstate basis  can be found in ref.\cite{Bertolini:1990if}. In
our analysis we take as an input $\Delta M_{B_d}^{\rm exp}=0.507
\pm 0.005\, ps^{-1} $, $\Delta M_{B_s}^{\rm exp}= 17.78 \pm 0.12\,
ps^{-1} $ \cite{TheHeavyFlavorAveragingGroup:2010qj} and $f_{B_s}=
231 \pm 15 MeV $\cite{Gamiz:2009ku},  $\hat{B}_q \simeq 1$
\cite{Gabbiani:1996hi}. For  constraining our parameter space
using ${B}_q-\bar{B}_q $ mixing, we require that $\Delta M_{B_q}$,
including SUSY contributions, should lie within the $2\sigma$
range of the experimental values.

Other relevant constraints on the SUSY parameter space  can be
obtained by requiring the radiative corrections to the CKM
elements do not exceed the experimental values
\cite{Crivellin:2008mq}, by studying the effect of a right-handed
coupling of quarks to the W-boson on the measurements of
$|V_{ub}|$ and $|V_{cb }|$~\cite{Crivellin:2009sd} and by applying
't~Hooft's naturalness criterion from the mass and CKM
renormalization \cite{Crivellin:2010gw}.

 Finally, we take into account one of the most restrictive  constraints that comes from
$ B_s\to \mu^+\mu^- $ decay. The analysis of this decay in the
context of the SM as well as NP models have been performed in the
literature in
refs.\cite{Skiba:1992mg,Choudhury:1998ze,Huang:2000sm,Bobeth:2001sq,Huang:2002ni,Chankowski:2003wz,Alok:2005ep,Blanke:2006ig,
Alok:2009wk,Buras:2010pi,Golowich:2011cx,Alok:2010zd,Alok:2011gv,Wang:2011aa}.
In the SM, this decay channel vanishes at tree level, while it
occurs at one-loop level with the charged gauge boson $W^\pm$ and
up-type quarks in the loop. In MSSM, $ B_s\to \mu^+\mu^- $ decay
can be generated at quark level via $ b \rightarrow s \mu^+ \mu^-$
transition at one-loop level.  The different  contributions to
this transition depend on the  particles propagated in the loop
namely, ($1$) Standard Model gauge boson $W^\pm$ and up-type
quarks (SM contribution); ($2$) charged Higgs $H^\pm$ and up-type
quarks (charged Higgs contribution); ($3$) chargino and scalar
up-type quarks (chargino contribution); ($4$) neutralino and
scalar down-type quarks (neutralino contribution); ($5$) gluino
and scalar down-type quarks (gluino
contribution)\cite{Huang:2002ni}. The branching ratio  $ B_s\to
\mu^+\mu^- $ including supersymmetric contributions is given  by
\cite{Huang:2002ni}

\begin{eqnarray}
{\rm BR}(B_s \rightarrow \mu^+ \mu^-) &=& \frac{G_F^2
\alpha^2_{\rm em}}{64 \pi^3} m^3_{B_s} \tau_{B_s} f^2_{B_s}
|\lambda^s_t|^2 \sqrt{1 - 4 \hat{m}^2}
[(1 - 4\hat{m}^2) |C_{Q_1}(m_b) - C^\prime_{Q_1}(m_b)|^2 + \nonumber\\
&& |C_{Q_2}(m_b) - C^\prime_{Q_2}(m_b) + 2\hat{m}( C_{10}(m_b) -
C^\prime_{10}(m_b) )|^2]\label{bsmumu}
\end{eqnarray}
where $\hat{m} = m_\mu/m_{B_s}$. In the SM, $C_{Q_1}=
C^\prime_{Q_1}= C_{Q_2}= C^\prime_{Q_2}= C^\prime_{10}=0$ and
\begin{eqnarray}
C_{10}=\frac{Y(x_t)}{\mbox{sin}^2\theta_W}.
\end{eqnarray}
where the loop function $Y(x_t)$ can be found in ref.
\cite{Altmannshofer:2009ne}.  For SUSY case, the complete
expressions for the Wilson coefficients $C_{Q_{1,2}}(m_W)$,
$C^\prime_{Q_{1,2}}(m_W)$, $C_{10}(m_W)$ and $C^\prime_{10}(m_W)$
in mass eigenstate basis can be found in Appendix A of
ref.\cite{Huang:2002ni}. The running of the Wilson coefficients
$C_{10}$ and $C_{Q_i}$ from $m_W$ to $m_b$ in the leading order
approximation (LO) is given in refs.\cite{operator,dhh}. The
evolution of part of the primed operators has been given in
ref.~\cite{bghw}.  The  SM prediction for the branching ratio  $
B_s\to \mu^+\mu^- $is $(3.2\pm0.2)\times10^{-9}$
\cite{Buras:2010mh}. The $B_s \to \mu^+ \mu^-$ decay has been
searched for at the Tevatron and the LHC. The CDF experiment has
reported and excess of events corresponding to a branching
fraction of
($1.8^{1.1}_{-0.9}$)$\times$10$^{-8}$~\cite{Aaltonen:2011fi}. The
LHCb and CMS collaborations did not observe any significant excess
and released a 95\% C.L.\ combined limit of
$\mathrm{BR}(B_s\to\mu^+\mu^-) < 1.1 \times
10^{-8}$~\cite{Chatrchyan:2011kr,arXiv:1112.0511,LHCBdata,CMSLHCB},
which is only $\sim$4 times above the SM predictions. Recently the
LHCb collaborators announced  a new upper limits  on the branching
ratios of $B_s\to\mu^+\mu^- $ and $B_d\to\mu^+\mu^- $ to be
$\mathrm{BR}(B_s\to\mu^+\mu^-) < 4.5 \times 10^{-9}$ and
$\mathrm{BR}(B_d\to\mu^+\mu^-) < 1.0 \times 10^{-9}$ at 95\% C.L.
In ref.\cite{Huang:2002ni}, it was shown that within non minimal
flavor violation MSSM and in the case of large $\tan\beta$,
BR($B_s \rightarrow \mu^+ \mu^-$) can be enhanced by a factor of
$10^3$ compared to SM prediction. In our analysis, we consider
also the constraint from $B_d \rightarrow \mu^+ \mu^-$ direct
search where the corresponding branching ratio can be obtained
easily from  BR($B_s \rightarrow \mu^+ \mu^-$) by the replacement
$ s\leftrightarrow d $ everywhere.  In our numerical analysis we
use the most recent limits $\mathrm{BR}(B_s\to\mu^+\mu^-) < 4.5
\times 10^{-9}$ and $\mathrm{BR}(B_d\to\mu^+\mu^-) < 1.0 \times
10^{-9}$.

 After scanning over the MSSM parameter space and imposing all
the above criteria, we find that gluino contributions are much
smaller than chargino contributions in agreement with
ref.\cite{Hofer:2010ee}. Thus in our discussion we will focus on
chargino contribution only although we include all contributions
in our results.

\begin{figure}[tbhp]
\includegraphics[width=6.5cm,height=7cm]{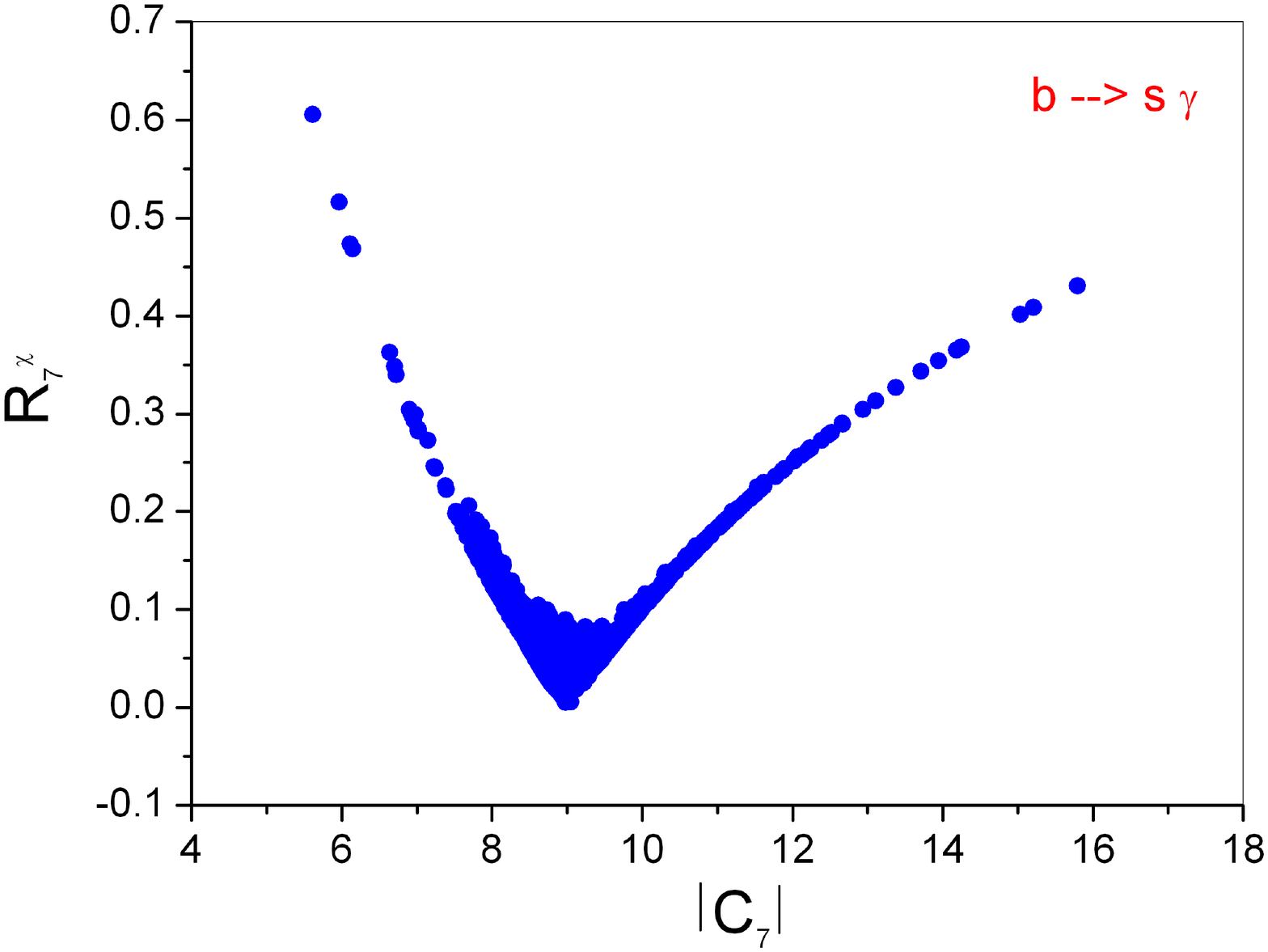}
\hspace{1.cm}
\includegraphics*[width=6.5cm,height=7cm]{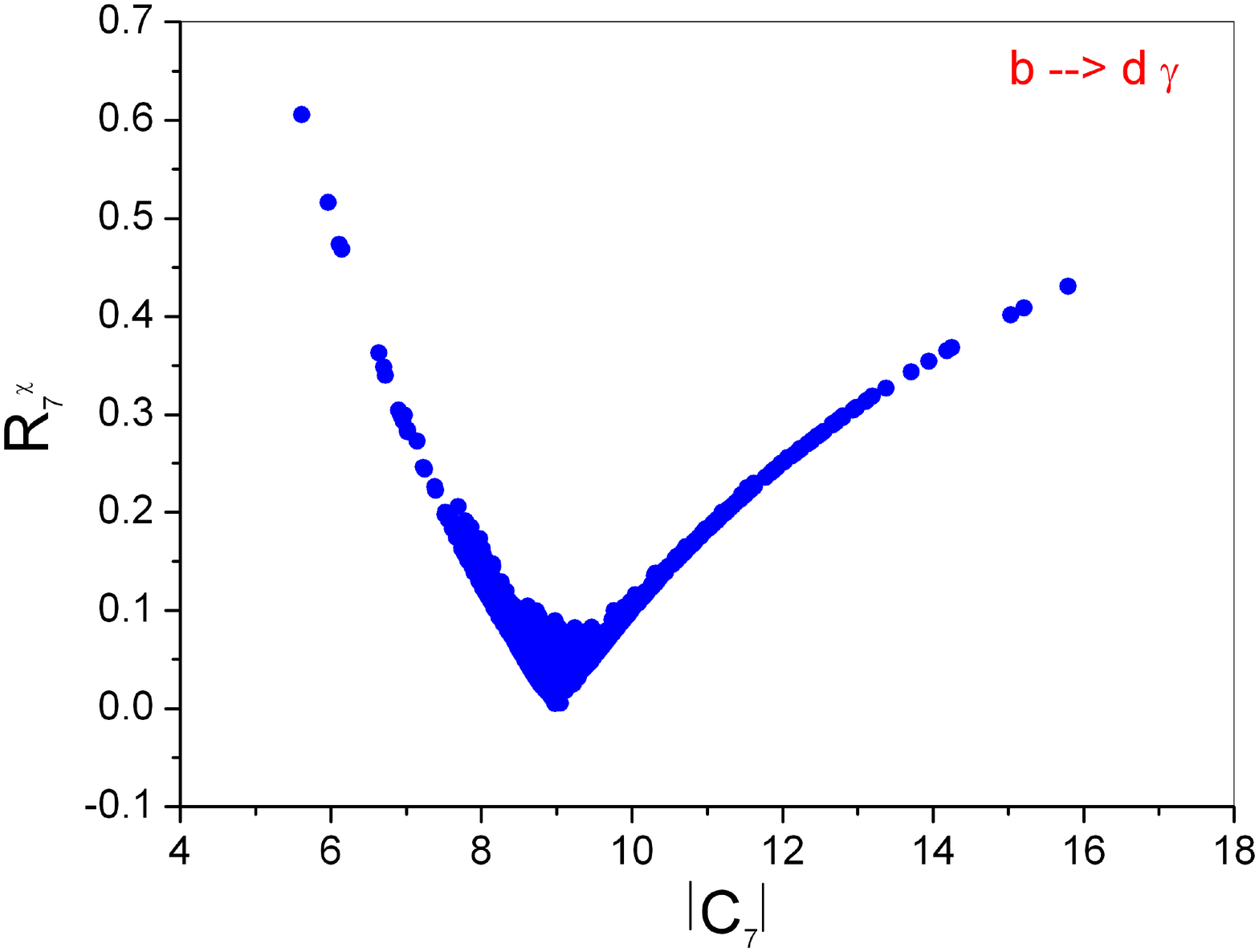}
\medskip
\caption{ $R^{\chi}_7 $ versus $|C_7|$ in units of $10^{-5}$. Left
diagram corresponds to points in the MSSM parameter space passing
the constraint from $b\to s\gamma$. Right diagram corresponds  to
the points passing the constraint from $b\to d\gamma$.}
\label{singlemas1}
\end{figure}

\begin{figure}[tbhp]
\includegraphics[width=6.5cm,height=7cm]{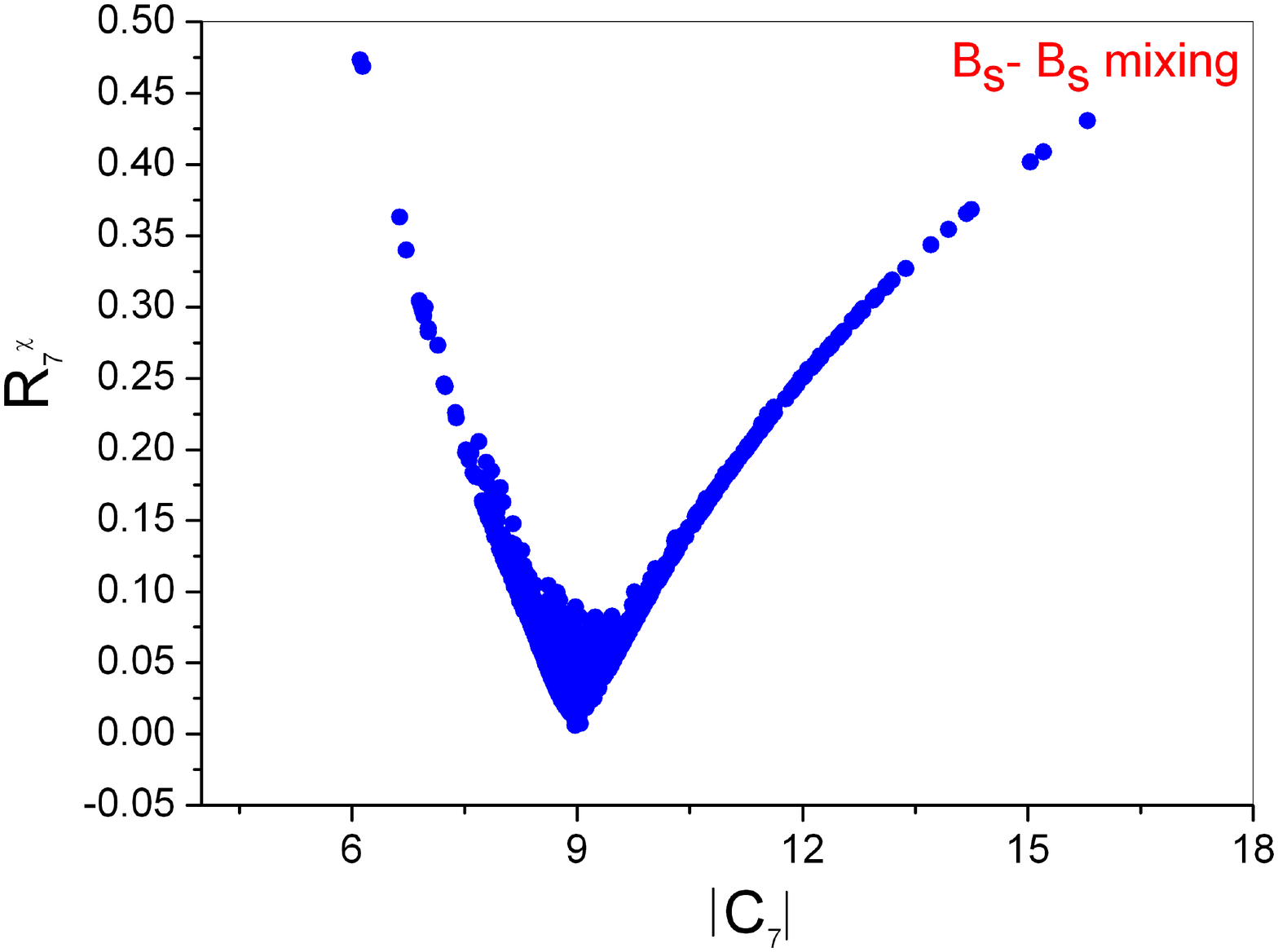}
\hspace{1.cm}
\includegraphics*[width=6.5cm,height=7cm]{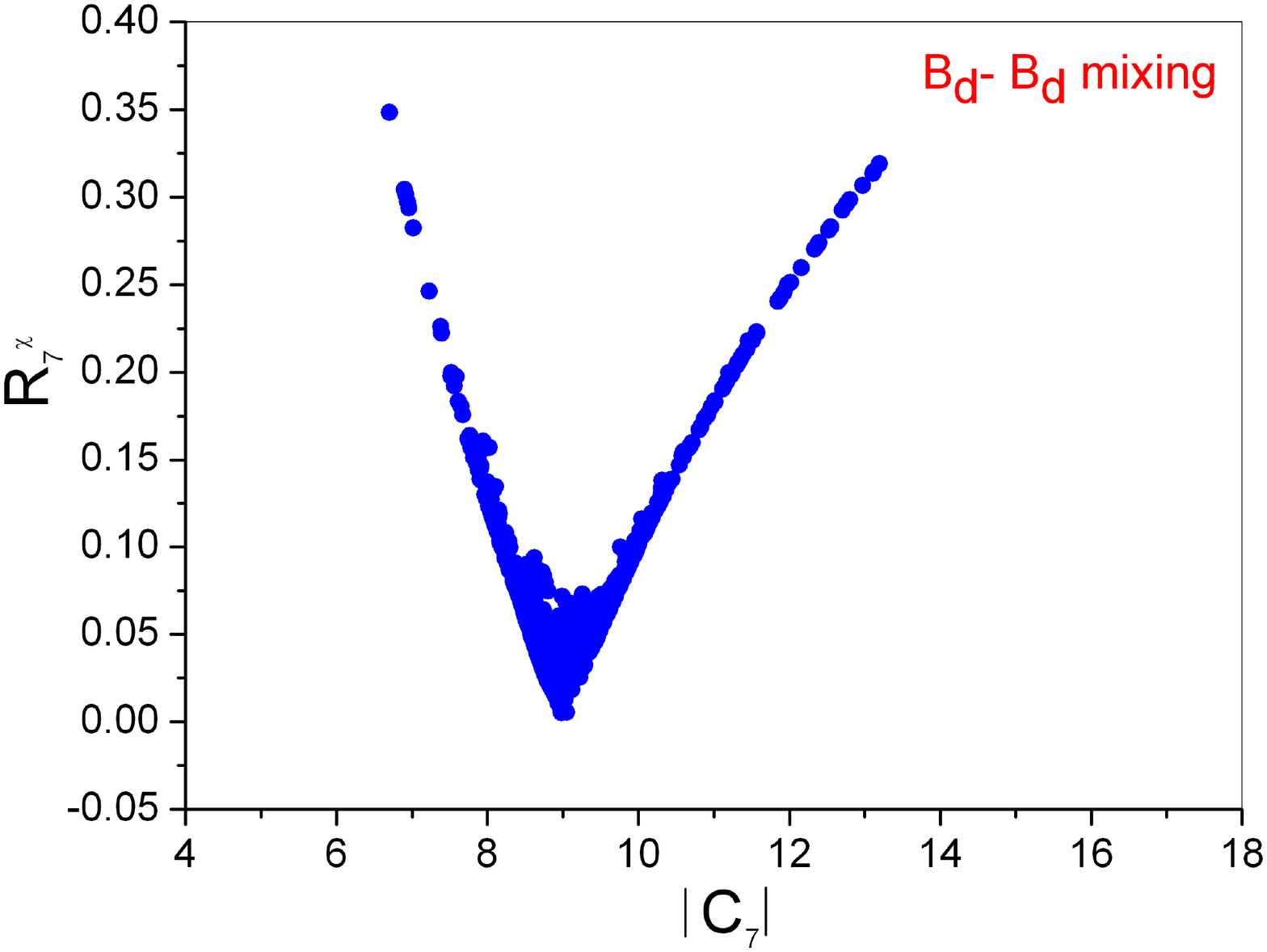}
\medskip
\caption{$R^{\chi}_7 $ versus $|C_7|$ in units of $10^{-5}$. Left
diagram corresponds to the points in the MSSM parameter space
passing the constraints from $B_s-B_s$ mixing. Right diagram
corresponds to the points  passing the constraint from
$B_d-B_d$mixing.} \label{singlemas2}
\end{figure}

\begin{figure}[tbhp]
\includegraphics[width=6.5cm,height=7cm]{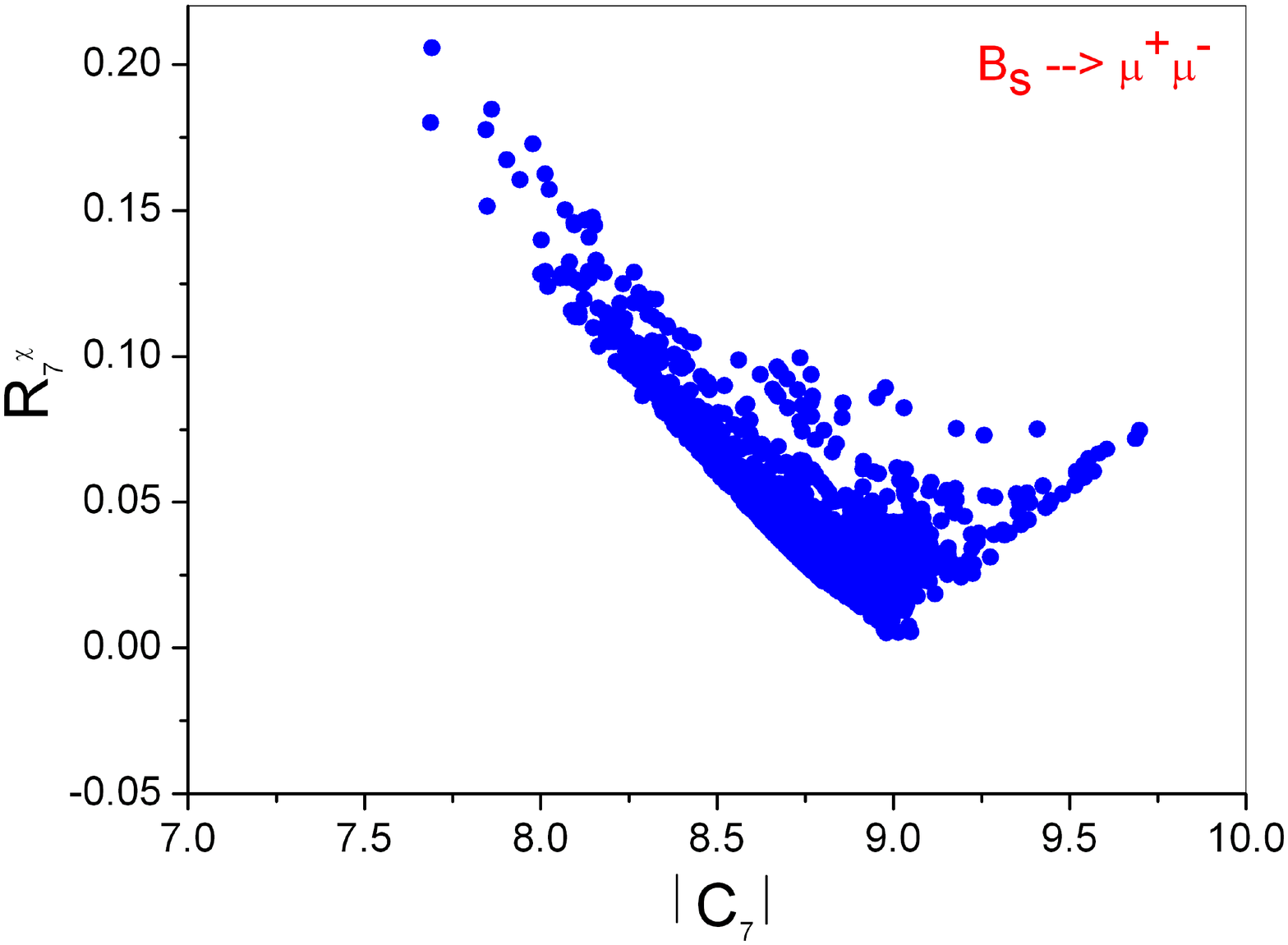}
\hspace{1.cm}
\includegraphics*[width=6.5cm,height=7cm]{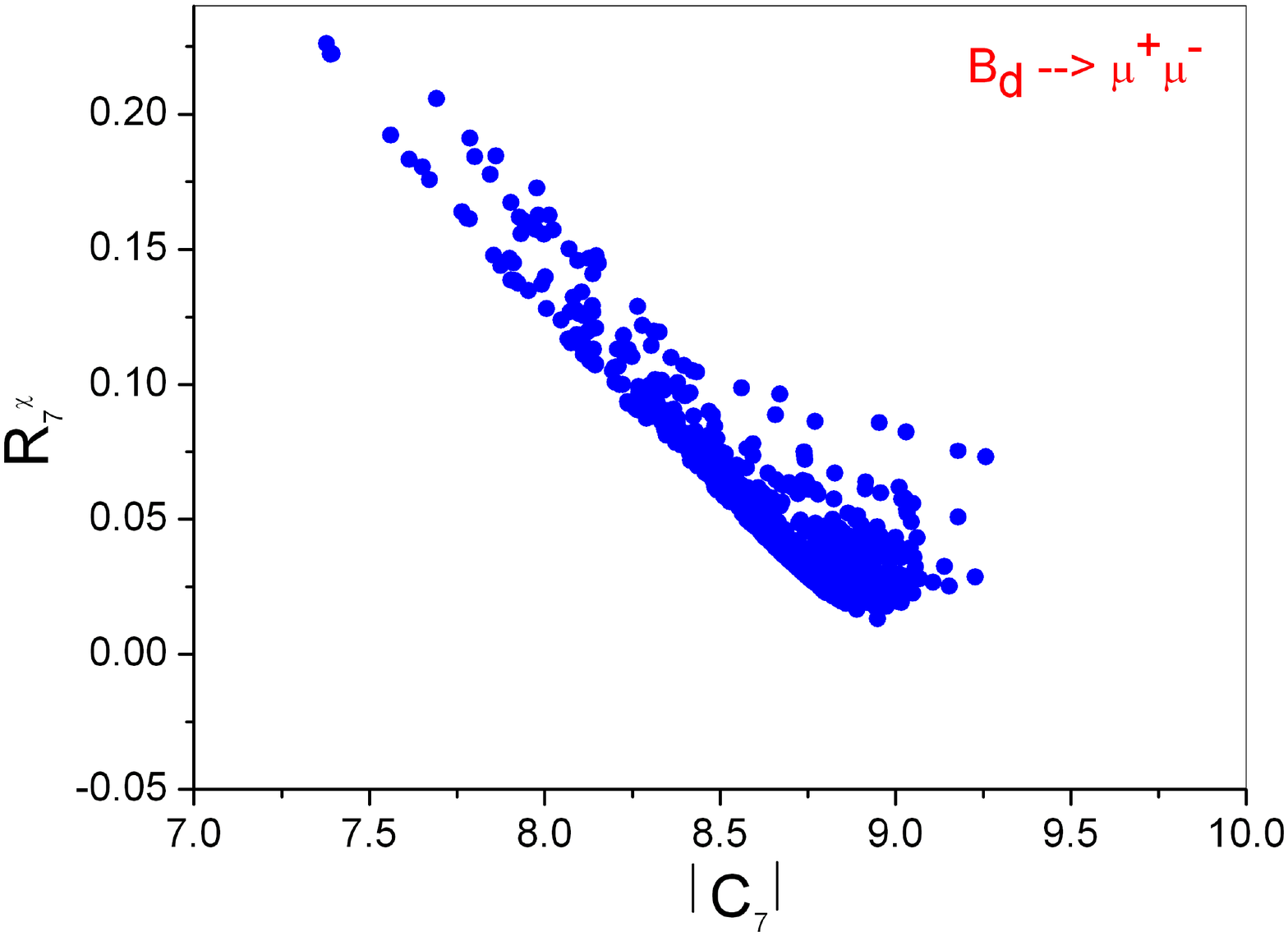}
\medskip
\caption{ $R^{\chi}_7 $ versus $|C_7|$ in units of $10^{-5}$. Left
diagram corresponds to the points in the MSSM parameter space
passing the constraints from $B_s\to \mu^+\mu^-$. Right diagram
corresponds to the points passing the constraint from $B_d\to
\mu^+\mu^-$.} \label{singlemas3}
\end{figure}

\begin{figure}[tbhp]
\includegraphics[width=6.5cm,height=7cm]{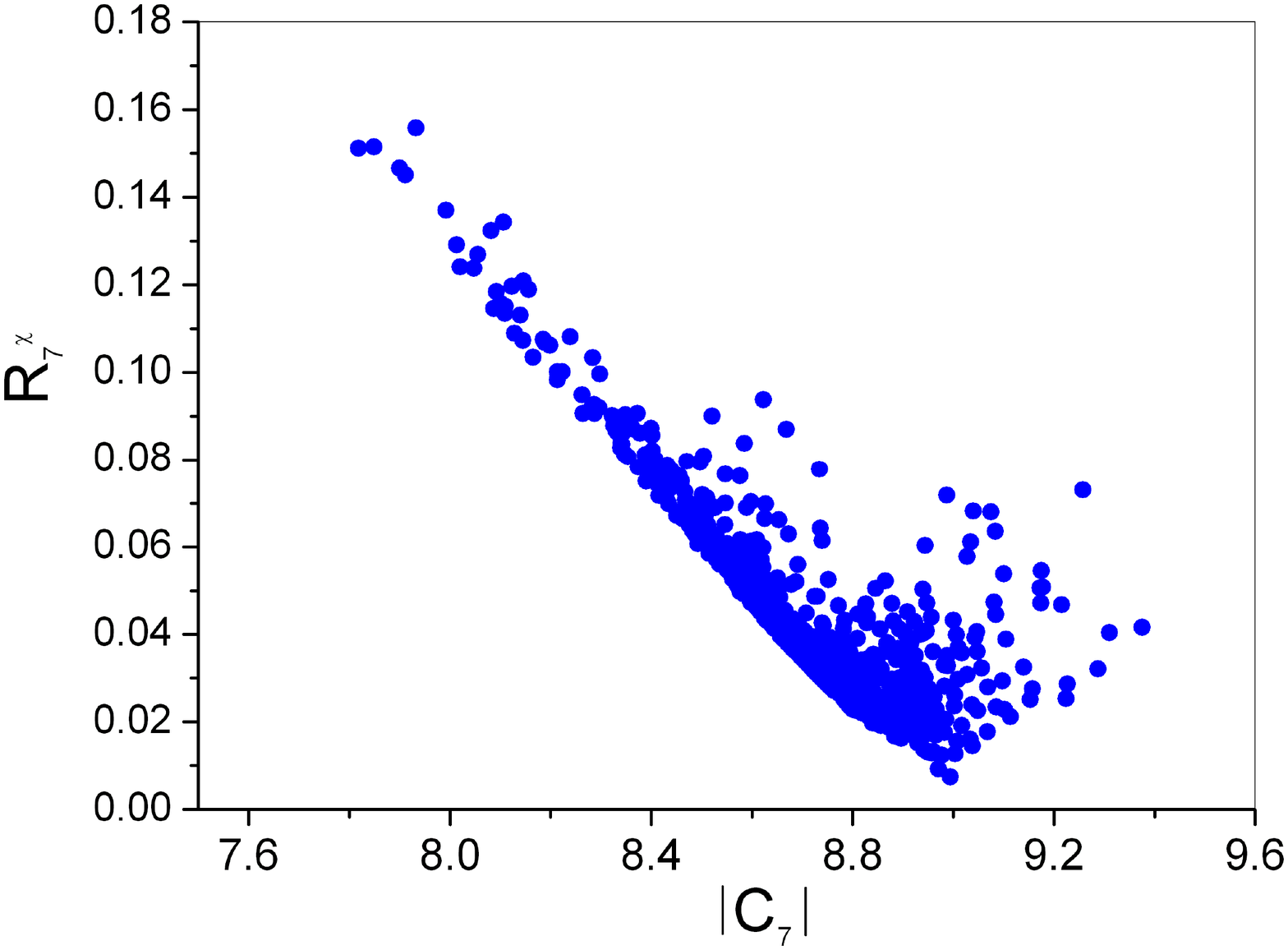}
\hspace{1.cm}
\includegraphics*[width=6.5cm,height=7cm]{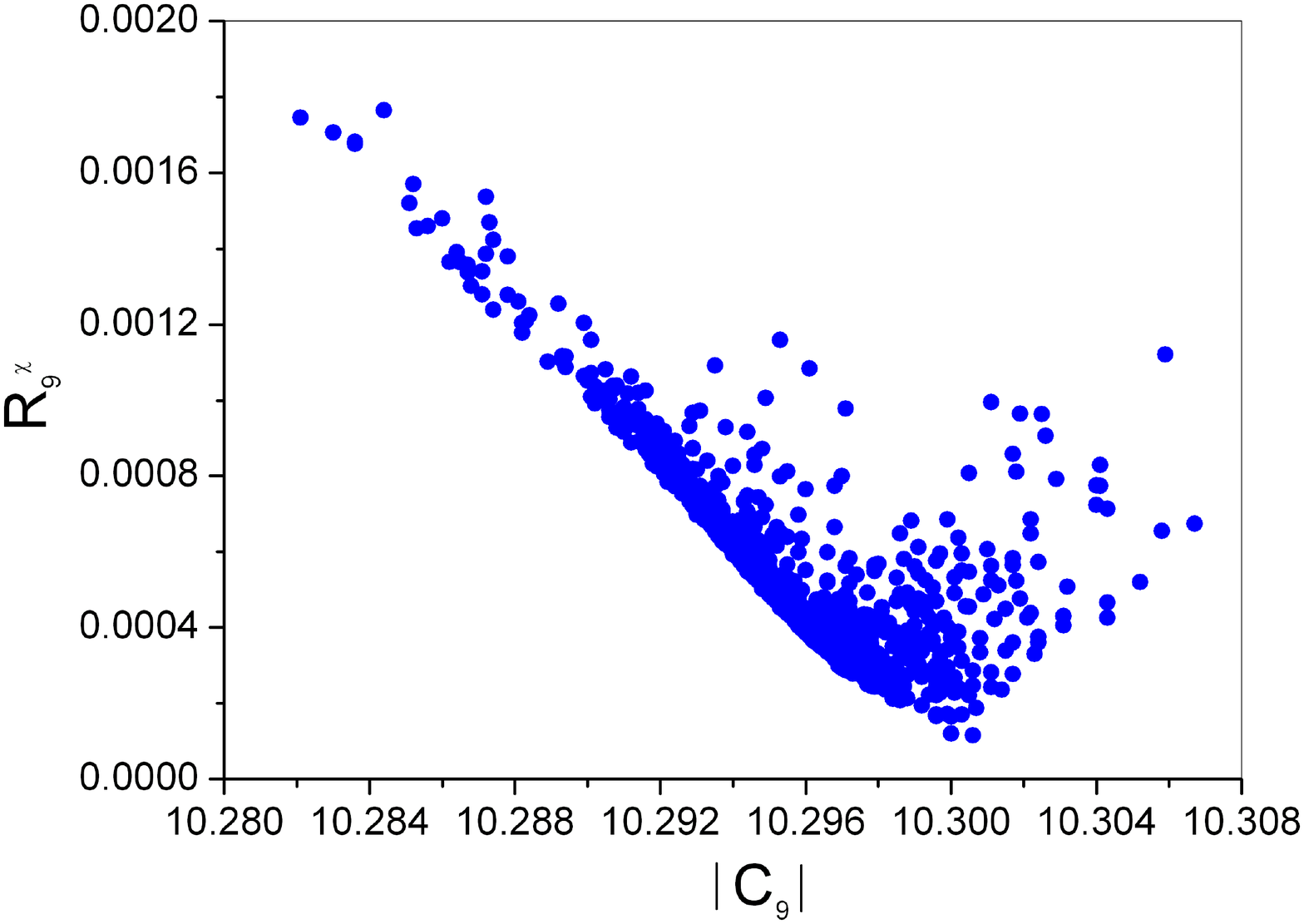}
\medskip
\caption{Left diagram corresponds to $R^{\chi}_7 $ versus $|C_7|$
in units of $10^{-5}$ while  right diagram corresponds to
$R^{\chi}_9 $ versus $|C_9|$ in units of $10^{-3}$ for points in
the MSSM parameter space passing all constraints discussed in the
text.} \label{singlemas4}
\end{figure}

In order to estimate the enhancement in the full Wilson
coefficients $ C_7 $ and $ C_9 $ due to chargino contribution we
define the two ratios: $R^{\chi}_7 = \frac{|C^{{\chi}}_7|}{|C_7|}$
and $R^{\chi}_9 = \frac{|C^{{\chi}}_9|}{|C_9|}$ where $C_7$ and
$C_9$ are the total Wilson coefficients. We start our numerical
analysis by displaying the results of imposing the different
constraints on the MSSM parameter space. For simplicity we only
show the plots of $R^{\chi}_7 $ versus $|C_7|$ for the points in
the MSSM parameter space passing one constraint per time.
Moreover, we show the corresponding constraints in the $b\to s $
and $b\to d $ transitions. This will help us to compare the
strength of the corresponding constraints in the two sectors. In
Fig.(\ref{singlemas1}) we plot $R^{\chi}_7 $ verses $|C_7|$ for
the  points allowed by $b\to s \gamma$ and $b\to d \gamma$
constraints. The left diagram corresponds to the points passing
$b\to s \gamma$ constraint and the right one corresponds to the
points passing  $b\to d \gamma$ constraint. For the points with
$C^{{\chi}}_7$ is much smaller than $C^{SM}_7$ we expect that
$R^{\chi}_7$ to be close to zero and  $|C_7|$ to be  close to
$|C^{SM}_7|\simeq 9 \times 10^{-5}$ which is clear from the kink
in Fig.(\ref{singlemas1}). On the other hand the points in the
parameter space that lead to  sign $C^{{\chi}}_7$
similar(opposite) to sign $C^{SM}_7$ will enhance(reduce) $|C_7|$
which in turn reduces(enhances)$R^{\chi}_7$  which can be seen in
the Figure. We see also from Fig.(\ref{singlemas1}) that the
maximum value of $R^{\chi}_7 $ is about 0.6 which means that $C_7$
can be enhanced by about 60\% for all points passing both
constraints. Clearly this indicates that these constraints are not
the strongest ones as we will see below. Next we consider the
constraints from $ B_q-B_q $ mixing where as before $q=d,s$. We
plot the corresponding graphs in Fig.(\ref{singlemas2}). The plots
in that figure have several features like the plots in
Fig.(\ref{singlemas1}). The differences between the two figures
are that $C_7$ can be enhanced by about 45\% and 35\%  for the
points passing $ B_s-B_s $ and $ B_d-B_d $ mixing constraints
respectively.  This implies that the constraints from $ B_q-B_q $
mixing are stronger than those from $b\to s \gamma$ and $b\to d
\gamma$ constraints. Moreover, the constraint from $ B_d-B_d $ is
stronger than that of $ B_s-B_s $ mixing constraint.  In
Fig.(\ref{singlemas3}) we display the points passing  the $ B_q\to
\mu^+\mu^- $ constraints where $q=d,s$. As can be seen from
Fig.(\ref{singlemas3}), $C_7$ can be enhanced by less than 20\% in
both plots.  Clearly $ B_q\to \mu^+\mu^- $ provides the strongest
constraints in  both $b\to s$ and $b\to d$ transitions. We notice
also from Fig.(\ref{singlemas3}) that the constraint from $ B_d\to
\mu^+\mu^-$ are slightly stronger than that of $ B_s\to \mu^+\mu^-
$ which is clear from  the maximum value that $|C_7|$ can reach
where we find that $|C_7|$ can reach $9.25\times 10^{-5}$ and
$9.75\times 10^{-5}$ after considering $ B_d\to \mu^+\mu^-$ and $
B_s\to \mu^+\mu^- $ respectively.

We now apply all constraints on the MSSM parameter space at the
same time and present our predictions for the ratios
$R^{\chi}_{7,9}$ and the branching ratios (BR) of $(\bar{B}_s\to
\phi \pi^0)$ and $(\bar{B}_s\to \phi \rho^0)$. In
Fig.(\ref{singlemas4}) we plot $R^{\chi}_7 $ verses $|C_7|$ and
$R^{\chi}_9 $  versus $|C_9|$ for the allowed points in the
parameter space. The left diagram corresponds to $R^{\chi}_7$
versus $|C_7|$ in units of $10^{-5}$ while the right diagram
corresponds to $R^{\chi}_9 $ versus $|C_9|$ in units of $10^{-3}$.
As can be seen from Fig.(\ref{singlemas4}) that, $C_7$ can be
enhanced by less than 20 \% and $C_9 $ can be enhanced by  less
than 1 \%. In order to explain this result we note that the
dominant contributions to $ C^{\tilde{\chi}}_7 $ and $
C^{\tilde{\chi}}_9 $ come from Z penguins which are expressed in
terms of $Y_{1Z}$ and $Y_{2Z}$ respectively as given in the
Appendix. The ratio  $ \mid Y_{2Z} / Y_{1Z} \mid \simeq 1.2 $
which means that the full Wilson coefficients $C_7$ and  $C_9$
will be enhanced by the same order of magnitude and since
$C^{SM}_9 \gg C^{SM}_7$ we can expect that $R^{\chi}_9 \ll
R^{\chi}_7 $ which is the same result that we obtained  in
Fig.(\ref{singlemas4}). As a consequence we expect that SUSY
contributions will not enhance the branching ratios of
$(\bar{B}_s\to \phi \pi^0)$ and $(\bar{B}_s\to \phi \rho^0)$
sizeably.

\begin{figure}[tbhp]
\includegraphics[width=6.5cm,height=7cm]{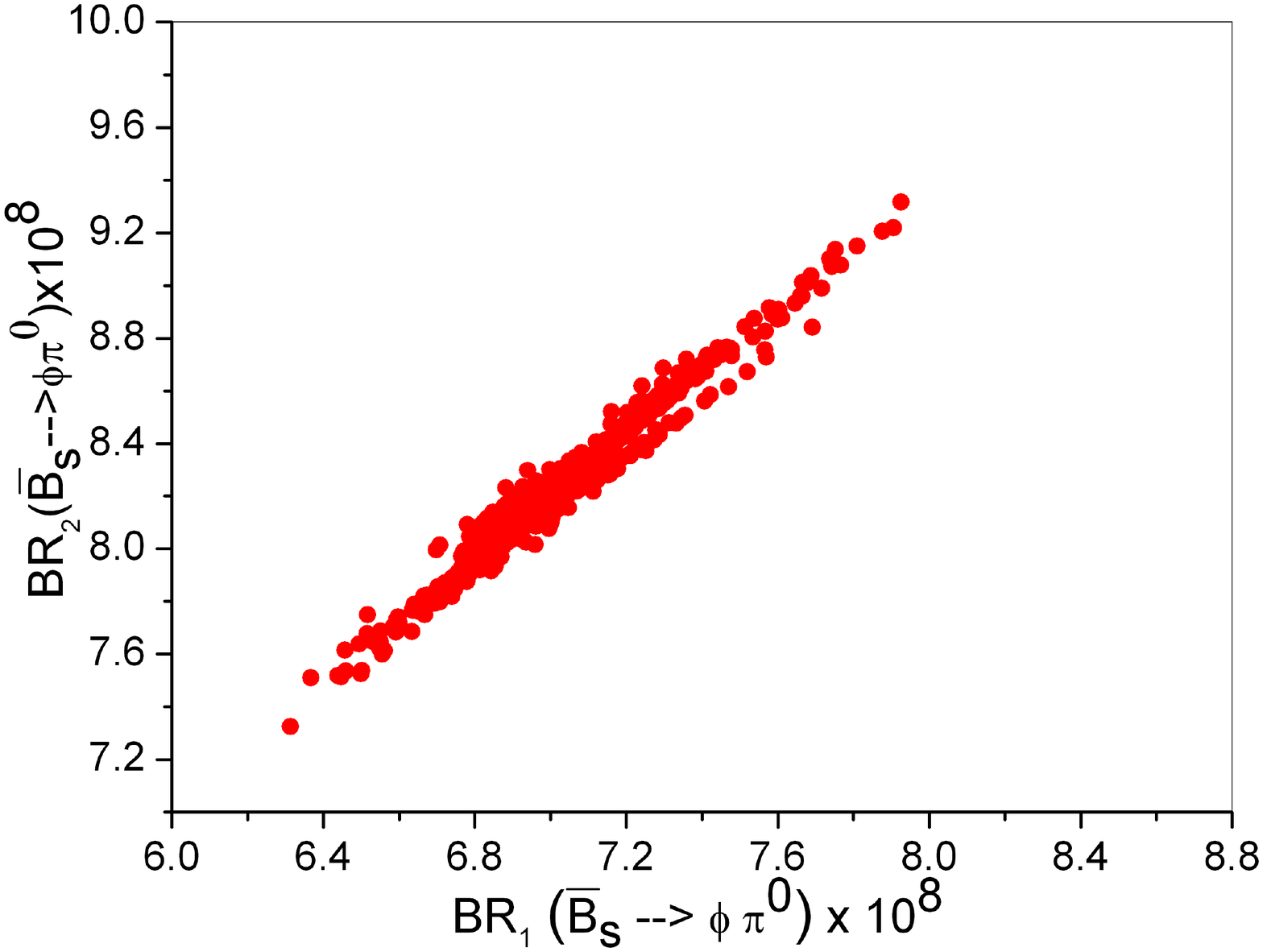}
\hspace{1.cm}
\includegraphics*[width=6.5cm,height=7cm]{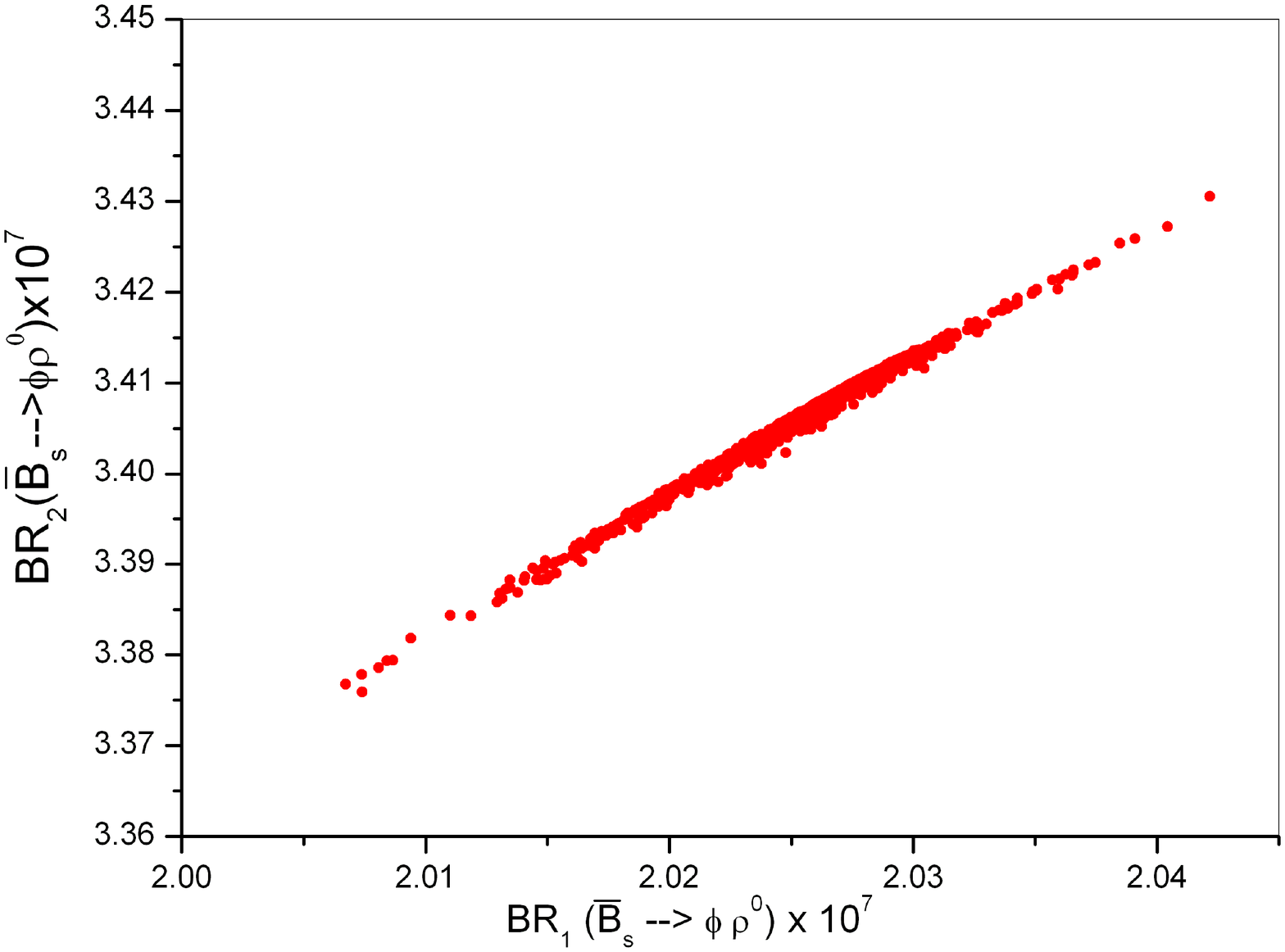}
\medskip
\caption{Branching ratios of $(\bar{B}_s\to \phi \pi^0)$ and
$(\bar{B}_s\to \phi \rho^0)$. Left plot corresponds to BR
$(\bar{B}_s\to \phi \pi^0) \times 10^8$ while right plot
corresponds to BR $(\bar{B}_s\to \phi \rho^0) \times 10^8$. In the
Figure $BR_1$ and $BR_2$ refer to the branching ratios correspond
to predictions 1 and 2 in Table~\ref{branch}.} \label{branchg1}
\end{figure}

 In Fig.(\ref{branchg1}) we plot the branching ratios  of
$(\bar{B}_s\to \phi \pi^0)$ and $(\bar{B}_s\to \phi \rho^0)$
resulting after the scan over the MSSM parameter space subjected
to the constraints discussed above. The left diagram corresponds
to BR $(\bar{B}_s\to \phi \pi^0)$ while the right diagram
corresponds to BR $(\bar{B}_s\to \phi \rho^0)$. In both diagrams
we present the predictions corresponding to solution $ 1 $ and $ 2
$ of the SCET parameters. Clearly from Fig.(\ref{branchg1})  SUSY
contributions can enhance BR $(\bar{B}_s\to \phi \pi^0)$ by about
14\% and  3\% with respect to the SM prediction for solution 1 and
2 of the SCET parameters respectively. For BR $(\bar{B}_s\to \phi
\rho^0)$, we see from Fig.(\ref{branchg1})  that SUSY
contributions  enhance  the branching ratios by about 1\% with
respect to the SM prediction for both solution 1 and 2 of the SCET
parameters.  The reason for the significant enhancement of the
branching ratios in case of $(\bar{B}_s\to \phi \pi^0)$ compared
to the branching ratios of $(\bar{B}_s\to \phi \rho^0)$ due to
SUSY contributions can be attributed to the difference  in the
sign of $C_7$ in their amplitudes given in
eqs.(\ref{pi1},\ref{rho1}) and eqs.(\ref{pi2},\ref{rho2}) for
solution 1 and  2  of the SCET parameters respectively. Thus an
enhancement of $C_7$ will lead to opposite  effects in the total
amplitudes of $(\bar{B}_s\to \phi \pi^0)$ and $(\bar{B}_s\to \phi
\rho^0)$ and thus in their branching ratios as we obtained in
Fig.(\ref{branchg1}).

We end this section by comparing our results with the results
given in ref.\cite{Hofer:2010ee}. We find that  no sizeable
enhancement of the EW penguins Wilson coefficients in the MSSM
with flavour-violation in the up sector that may lead to large
effects in the decays $ \bar{B}_s\to \phi \pi^0$ and $
\bar{B}_s\to \phi \rho^0 $ in agreement with the conclusion of
ref.\cite{Hofer:2010ee}.

\section{Conclusion}\label{sec:conclusion}

In this article we have studied the decay modes $\bar{B}_s\to \phi
\pi^0$ and $\bar{B}_s\to \phi \rho^0$ using SCET. Within SM, we
find that  BR $\bar{B}_s\to \phi \pi^0 = 7_{-1-2}^{+1+2} \times
10^{-8} $ and BR $\bar{B}_s\to \phi \pi^0 =9_{-1-4}^{+1+3}\times
10^{-8} $ corresponding to
 solution $1$ and solution $2$ of the SCET parameters respectively. In addition
we find that  BR  $\bar{B}_s\to \phi \rho^0 =
20.2^{+1+9}_{-1-12}\times 10^{-8} $  and BR $ \bar{B}_s\to \phi
\rho^0 = 34.0^{+1.5 + 15}_{-1.5-22}\times 10^{-8} $ corresponding
to solution $1$ and solution $2$ of the SCET parameters
respectively.  Clearly, within SM, the decay modes $\bar{B}_s\to
\phi \pi^0$ and $\bar{B}_s\to \phi \rho^0$ have tiny branching
ratios of order $\sim 10^{-7}$ leading to a difficulty in
observing them. As a consequence any significant enhancement of
their branching ratios making them observable  at LHC will be a
clear indication of New Physics beyond SM.

 We have analyzed SUSY contributions to the branching ratios of
$(\bar{B}_s\to \phi \pi^0)$ and $(\bar{B}_s\to \phi \rho^0)$
decays using SCET. We have adopted in our analysis exact
diagonalization of the squark mass matrices. We have shown that,
BR $(\bar{B}_s\to \phi \pi^0)$ can be enhanced by about 14\% and
3\% with respect to the SM predictions for solution 1 and 2 of the
SCET parameters respectively. For BR $(\bar{B}_s\to \phi \rho^0)$,
we find that  BR $(\bar{B}_s\to \phi \rho^0)$ is enhanced  by
about 1\% with respect to the SM predictions for both solution 1
and 2 of the SCET parameters.  Clearly, SUSY contributions
obtained from gluino and chargino mediation can not lead to a
significant enhancement of the branching ratios of $(\bar{B}_s\to
\phi \pi^0)$ and $(\bar{B}_s\to \phi \rho^0)$ decays  making them
easily detectable  at LHC. Moreover, in case of observation of
these decays due to  improved experimental techniques it will not
be possible to pin down the SUSY contributions.

\section*{Acknowledgement}

Gaber Faisel's  work is supported by the National Science Council
of R.O.C. under grants NSC 99-2112-M-008-003-MY3 and NSC
100-2811-M-008-036.
\appendix
\section{Appendix}

The chargino contributions to the Wilson coefficients are given
by\cite{Du:1997zc,Kruger:2000ff}
\begin{eqnarray}
C_1^{{\chi^{}}}&=&0,\nonumber\\
C_2^{{\chi^{}}}&=&0,\nonumber\\
C_3^{{\chi^{}}}&=& -\frac{\alpha_s}{24\pi}Z,\nonumber\\
C_4^{{\chi^{}}}&=& \frac{\alpha_s}{8\pi}Z,\nonumber\\
C_5^{{\chi^{}}}&=& -\frac{\alpha_s}{24\pi}Z,\nonumber\\
C_6^{{\chi^{}}}&=& \frac{\alpha_s}{8\pi}Z,\nonumber\\
C_7^{{\chi^{}}}&=&\frac{\alpha}{6\pi}Y_1,\nonumber\\
C_8^{{\chi^{}}}&=&0,\nonumber\\
C_9^{{\chi^{}}}&=&\frac{\alpha}{6\pi}Y_2,\nonumber\\
C_{10}^{{\chi^{}}}&=&0.\nonumber\\
C_{7\gamma}^{{\chi^{}}}&=& -\frac{1}{6g^2
}\sum_{A=1}^6\sum_{j=1}^2
\frac{M_W^2}{m^2_{\tilde{\chi_j}^{\pm}}}\Bigg[(X_j^{U_L\dagger})_{2A}(X_j^{U_L})_{A3}
f_1(\frac{m^2_{\tilde{u}_A}}{m^2_{\tilde{\chi}_j^{\pm}}}) -2
(X_j^{U_L\dagger})_{2A}(X_j^{U_R})_{A3}\frac{m_{\tilde{\chi_j}^{\pm}}}{m_b}
f_2(\frac{m^2_{\tilde{u}_A}}{m^2_{\tilde{\chi}_j^{\pm}}})\Bigg],\nonumber\\
C_{8g}^{{\chi^{}}}&=&-\frac{1}{6g^2 }\sum_{A=1}^6\sum_{j=1}^2
\frac{M_W^2}{m^2_{\tilde{\chi_j}^{\pm}}}\Bigg[(X_j^{U_L\dagger})_{2A}(X_j^{U_L})_{A3}
g_1(\frac{m^2_{\tilde{u}_A}}{m^2_{\tilde{\chi}_j^{\pm}}}) -2
(X_j^{U_L\dagger})_{2A}(X_j^{U_R})_{A3}\frac{m^2_{\tilde{\chi_j}^{\pm}}}{m_b}
g_2(\frac{m^2_{\tilde{u}_A}}{m^2_{\tilde{\chi}_j^{\pm}}})\Bigg]
\end{eqnarray}

 We can write $Y_i=Y_{iZ}+Y_{i\gamma}$ for $i=1,2$  where
$Y_{iZ}$ refers to the contribution from chargino loops with a
Z-boson coupling to the quark pair ,  $Y_{i\gamma}$ refers to
contribution from chargino loops with a photon coupling to the
quark pair and Z refers to contribution from chargino loops with a
gluon coupling to the quark pair. Their explicit expressions are
given as

\begin{eqnarray}
Z&=&-\displaystyle\frac{1}{3g^2_2V^*_{ts}V_{tb}}\displaystyle\sum_{A=1}^6
     \displaystyle\sum_{i=1}^2\displaystyle\frac{m^2_W}{m^2_{\tilde{u}_A}}
     (X_i^{U_L})_{2A}^+(X_i^{U_L})_{A3}
     f_5\left(\displaystyle\frac{m^2_{\tilde{\chi}_i^{\pm}}}
     {m^2_{\tilde{u}_A}}\right),\nonumber\\
 Y_{1Z}&=&-\displaystyle\frac{2}{g^2_2V^*_{ts}V_{tb}}\displaystyle\sum_{A,B=1}^6
       \displaystyle\sum_{i,j=1}^2(X_i^{U_L})_{2A}^+(X_j^{U_L})_{B3}
    \left\{c_2(m^2_{\tilde{\chi}_i^{\pm}},m^2_{\tilde{u}_A},m^2_{\tilde{u}_B})
    (\Gamma^{U_L}\Gamma^{U_L+})_{AB}\delta_{ij}\right.\nonumber\\[4mm]
 &&-c_2(m^2_{\tilde{u}_A},m^2_{\tilde{\chi}_i^{\pm}},m^2_{\tilde{\chi}_j^{\pm}})
   \delta_{AB}V_{i1}^*V_{j1}+\displaystyle\frac{1}{2}m_{\tilde{\chi}_i^{\pm}}
   m_{\tilde{\chi}_j^{\pm}}
   c_0(m^2_{\tilde{u}_A},m^2_{\tilde{\chi}_i^{\pm}},m^2_{\tilde{\chi}_j^{\pm}})
   \delta_{AB}U_{i1}U_{j1}^*\left.\right\},\nonumber\\[4mm]
 Y_{2Z}&=&\left(\displaystyle\frac{1}{sin^2\theta_w}-2\right)
     \displaystyle\frac{1}{g^2_2V^*_{ts}V_{tb}}\displaystyle\sum_{A,B=1}^6
       \displaystyle\sum_{i,j=1}^2(X_i^{U_L})_{2A}^+(X_j^{U_L})_{B3}
    \left\{c_2(m^2_{\tilde{\chi}_i^{\pm}},m^2_{\tilde{u}_A},m^2_{\tilde{u}_B})
    (\Gamma^{U_L}\Gamma^{U_L+})_{AB}\delta_{ij}\right.\nonumber\\[4mm]
 &&-c_2(m^2_{\tilde{u}_A},m^2_{\tilde{\chi}_i^{\pm}},m^2_{\tilde{\chi}_j^{\pm}})
   \delta_{AB}V_{i1}^*V_{j1}+\displaystyle\frac{1}{2}m_{\tilde{\chi}_i^{\pm}}
   m_{\tilde{\chi}_j^{\pm}}
   c_0(m^2_{\tilde{u}_A},m^2_{\tilde{\chi}_i^{\pm}},m^2_{\tilde{\chi}_j^{\pm}})
   \delta_{AB}U_{i1}U_{j1}^*\left.\right\},\nonumber\\
Y_{1\gamma}&=&Y_{2\gamma}=-\displaystyle\frac{1}{9g^2_2V^*_{ts}V_{tb}}\displaystyle\sum_{A=1}^6
     \displaystyle\sum_{i=1}^2\displaystyle\frac{m^2_W}{m^2_{\tilde{u}_A}}
     (X_i^{U_L})_{2A}^+(X_i^{U_L})_{A3}
     f_4\left(\displaystyle\frac{m^2_{\tilde{\chi}_i^{\pm}}}
     {m^2_{\tilde{u}_A}}\right).
\end{eqnarray}
where
 \begin{eqnarray}  X_i^{UL} &=& g \bigg[
-V^*_{i1} \Gamma^{UL} + V^*_{i2} \Gamma^{UR}
\frac{M_U}{\sqrt{2}m_W \sin\beta} \bigg] K\nonumber\\
X_i^{UR} &=& g \bigg[ U_{i2} \Gamma^{UL} K \frac{M_D}{\sqrt{2} m_W
\cos\beta} \bigg]
\end{eqnarray}
 and the loop functions are given as follows

\begin{eqnarray}
g_1(x)&=& \frac{-6 - 15 x + 3 x^2}{6(1 - x)^3} + \frac{-3 x
lnx}{(1 - x)^4}\nonumber\\
 g_2(x)&=&\frac{3 x + 3}{2(1 - x)^2} + \frac{3 x lnx}{(1 -
x)^3} - \frac{3}{2},\nonumber\\
f_1(x)&=&\frac{-7 + 5 x + 8 x^2}{6(1 - x)^3}-\frac{x(2 - 3 x )lnx}{(1 - x)^4},\nonumber\\
f_2(x)&=&\frac{x(3 - 5 x )}{2(1 - x)^2}+\frac{ x((2 - 3
x) lnx)}{(1 - x)^3},\nonumber\\
f_4(x)&=&\displaystyle\frac{52-101x+43x^2}{6(1-x)^3}+
      \displaystyle\frac{6-9x+2x^3}{(1-x)^4}lnx,\nonumber\\
f_5(x)&=&\displaystyle\frac{2-7x+11x^2}{6(1-x)^3}+
         \displaystyle\frac{x^3}{(1-x)^4}lnx,\nonumber\\
c_0(m_1^2,m_2^2,m_3^2)&=&-\left(
     \displaystyle\frac{m_1^2ln(m_1^2/\mu^2)}{(m_1^2-m_2^2)
      (m_1^2-m_3^2)}+(m_1\leftrightarrow m_2)+(m_1\leftrightarrow m_3)\right),
       \nonumber\\
c_2(m_1^2,m_2^2,m_3^2)&=&\displaystyle\frac{3}{8}-
      \displaystyle\frac{1}{4}\left(
      \displaystyle\frac{m_1^4ln(m_1^2/\mu^2)}{(m_1^2-m_2^2)
      (m_1^2-m_3^2)}+(m_1\leftrightarrow m_2)+(m_1\leftrightarrow m_3)\right).
\end{eqnarray}

The gluino contributions to the Wilson coefficients  are  given
by\cite{Harnik:2002vs}
\begin{eqnarray}
C_1^{\tilde{g}}&=&0,\nonumber\\
C_2^{\tilde{g}}&=&0,\nonumber\\
C^{\tilde{g}}_3 &=& \frac{\als^2 \sqrt{2}}{m_{\tilde{g} }^2G_F}
\left( \sum_{AB} \Gamma^{L\ast}_{sA} \Gamma^{L}_{bA}
\Gamma^{L\ast}_{sB} \Gamma^{L}_{sB} \left[ -\frac{1}{9}
B_1(x_A,x_B)-\frac{5}{9}
B_2(x_A,x_B) \right] \right. \nonumber \\
&& \ \ \ \ \ \ \ \left. + \sum_A \Gamma^{R\ast}_{sA}
\Gamma^{R}_{bA}
\left[ -\frac{1}{18} C_1(x_A)+\frac{1}{2} C_2(x_A) \right] \right) \nonumber\\
C^{\tilde{g}}_4 &=& \frac{\als^2\sqrt{2}}{m_{\tilde{g}}^2G_F}
\left( \sum_{AB} \Gamma^{L\ast}_{sA} \Gamma^{L}_{bA}
\Gamma^{L\ast}_{sB} \Gamma^{L}_{sB} \left[ -\frac{7}{3}
B_1(x_A,x_B)+\frac{1}{3}
B_2(x_A,x_B) \right] \right. \nonumber \\
&& \ \ \ \ \ \ \ \left. + \sum_A \Gamma^{L\ast}_{sA}
\Gamma^{L}_{bA}
\left[ \frac{1}{6} C_1(x_A)-\frac{3}{2} C_2(x_A) \right] \right) \nonumber\\
 C^{\tilde{g}}_5 &=&
\frac{\als^2\sqrt{2}}{m_{\tilde{g}}^2G_F} \left( \sum_{AB}
\Gamma^{L\ast}_{sA} \Gamma^{L}_{bA} \Gamma^{R\ast}_{sB}
\Gamma^{R}_{sB} \left[ \frac{10}{9} B_1(x_A,x_B)+\frac{1}{18}
B_2(x_A,x_B) \right] \right. \nonumber \\
&& \ \ \ \ \ \ \ \left. + \sum_A \Gamma^{L\ast}_{sA}
\Gamma^{L}_{bA}
\left[ -\frac{1}{18} C_1(x_A)+\frac{1}{2} C_2(x_A) \right] \right) \nonumber\\
 C^{\tilde{g}}_6 &=&
\frac{\als^2\sqrt{2}}{m_{\tilde{g}}^2G_F} \left( \sum_{AB}
\Gamma^{L\ast}_{sA} \Gamma^{L}_{bA} \Gamma^{R\ast}_{sB}
\Gamma^{R}_{sB} \left[ -\frac{2}{3} B_1(x_A,x_B)+\frac{7}{6}
B_2(x_A,x_B) \right] \right. \nonumber \\
&& \ \ \ \ \ \ \ \left. + \sum_A \Gamma^{L\ast}_{sA}
\Gamma^{L}_{bA}
\left[ \frac{1}{6} C_1(x_A)-\frac{3}{2} C_2(x_A) \right] \right)\nonumber \\
 C^{\tilde{g}}_{7\gamma} &=&
\frac{\als\pi\sqrt{2}}{m_{\tilde{g}}^2G_F} \left( \sum_A
\Gamma^{L\ast}_{sA} \Gamma^{L}_{bA} \left[ -\frac{4}{9}
D_1(x_A)\right] + \frac{m_{\tilde{g}}}{m_b} \sum_A
\Gamma^{L\ast}_{sA} \Gamma^{R}_{bA}
\left[ -\frac{4}{9} D_2(x_A)\right]\right)\nonumber \\
 C^{\tilde{g}}_{8g} &=&
\frac{\als\pi\sqrt{2}}{m_{\tilde{g}}^2G_F} \left( \sum_A
\Gamma^{L\ast}_{sA} \Gamma^{L}_{bA} \left[ -\frac{1}{6}
D_1(x_A)+\frac{3}{2} D_3(x_A) \right]
\right. \nonumber \\
&& \ \ \ \ \ \ \ \left. + \frac{m_{\tilde{g}}}{m_b} \sum_A
\Gamma^{L\ast}_{sA} \Gamma^{R}_{bA} \left[ -\frac{1}{6}
D_2(x_A)+\frac{3}{2} D_4(x_A) \right] \right).
\end{eqnarray}
 where $x_A\equiv m_{\tilde{d}_A}^2/m_{\tilde{g}}^2$ and the loop
functions are given by
\begin{eqnarray}
    B_1(x_A,x_B) &=& -\frac{x_A^2 \log x_A}{4(x_A-x_B)(x_A-1)^2}
    -\frac{x_B^2 \log x_B}{4(x_B-x_A)(x_B-1)^2} -\frac{1}{4(x_A-1)(x_B-1)}, \nonumber \\
    B_2(x_A,x_B) &=&  -\frac{x_A \log x_A}{(x_A-x_B)(x_A-1)^2}
    -\frac{x_B \log x_B}{(x_B-x_A)(x_B-1)^2} -\frac{1}{(x_A-1)(x_B-1)},\nonumber \\
    C_1 (x) &=&  \frac{2x^3-9x^2+18x-11-6\log x}{36(1-x)^4}, \nonumber \\
    C_2 (x) &=& \frac{-16x^3+45x^2-36x+7+6x^2(2x-3)\log x}{36(1-x)^4},\nonumber \\
    D_1 (x) &=& \frac{-x^3+6x^2-3x-2-6x\log x}{6(1-x)^4},\nonumber \\
    D_2 (x) &=& \frac{-x^2+1+2x\log x}{(x-1)^3},\nonumber \\
    D_3 (x) &=&  \frac{2x^3+3x^2-6x+1-6x^2\log x}{6(1-x)^4},\nonumber \\
    D_4 (x) &=& \frac{-3x^2+4x-1+2x^2\log x}{(x-1)^3}.
\end{eqnarray}

The Wilson coefficients $C^{\tilde{g}}_i$ for $i=7,...,10$ are
given by \cite{Du:1997zc}

 \bea
C_7^{\tilde{g}}&=&\frac{\alpha}{6\pi}Y_1,\nonumber\\
C_8^{\tilde{g}}&=&0,\nonumber\\
C_9^{\tilde{g}}&=&\frac{\alpha}{6\pi}Y_2,\nonumber\\
C_{10}^{\tilde{g}}&=&0.\nonumber\\
\eea

where as before  we  write $Y_i=Y_{iZ}+Y_{i\gamma}$ for $i=1,2$
where $Y_{iZ}$ refers to the contribution from gluino loops with a
Z-boson coupling to the quark pair, $Y_{i\gamma}$ refers to
contribution from gluino loops with a photon coupling to the quark
pair.

\bea Y_{1Z}&=&-\displaystyle\frac{16g^2_3}{3g^2_2
V^*_{ts}V_{tb}}\displaystyle\sum_{A,B=1}^6
(\Gamma^{D_L})^{\dagger}_{2A}(\Gamma^{D_L})_{B3}
c_2(m^2_{\tilde{g}},m^2_{\tilde{d}_A},m^2_{\tilde{d}_B})(\Gamma^{D_R}\Gamma^{D_R{\dagger}})_{AB},\nonumber\\
Y_{2Z}&=&\displaystyle\frac{8g^2_3}{3g^2_2
V^*_{ts}V_{tb}}(\frac{1}{\sin^2\theta_w}-2)\displaystyle\sum_{A,B=1}^6
(\Gamma^{D_L})^{\dagger}_{2A}(\Gamma^{D_L})_{B3}
c_2(m^2_{\tilde{g}},m^2_{\tilde{d}_A},m^2_{\tilde{d}_B})(\Gamma^{D_R}\Gamma^{D_R{\dagger}})_{AB}
\eea

and

\bea
Y_{1\gamma}=Y_{2\gamma}&=&\displaystyle\frac{4g^2_3}{81g^2_2V^*_{ts}V_{tb}}\displaystyle\sum_{A=1}^6
\displaystyle\frac{m^2_W}{m^2_{\tilde{d}_A}}
(\Gamma^{D_L})^{\dagger}_{2A}(\Gamma^{D_L})_{A3}
f_6\left(\displaystyle\frac{m^2_{\tilde{g}}}
{m^2_{\tilde{d}_A}}\right),\nonumber\\
\eea

where \bea f_6(x)&=&\displaystyle\frac{2-7x+11x^2}{(1-x)^3}+
\displaystyle\frac{6 x^3}{(1-x)^4}lnx,\nonumber\\
\eea

It should be noted that the expressions for $Y_i$  given in the
last two equations are consistent with the corresponding
expressions given in ref.\cite{Cho:1996we} after replacing the
lepton charge with the up quark charge and taking into account the
effective Hamiltonian conventions used in ref.\cite{Cho:1996we}.
We further correct $Y_{1\gamma}$ and $ Y_{2\gamma}$ given in
equation B10 in ref.\cite{Du:1997zc}. We also neglect the small
contributions from the box diagrams to $C_9$ as the dominant
contributions is due to Z penguins\cite{Hofer:2010ee}.

\end{document}